\documentclass[12pt,preprint]{aastex}
\usepackage[scaled=0.9]{helvet}

\begin{document}

\noindent\textbf{A candidate redshift $z\approx 10$ galaxy and rapid changes in that population at an age of 500 Myr}


R. J. Bouwens$^{1,2}$, G. D. Illingworth$^{1}$, I. Labbe$^{3}$, P. A. Oesch$^{4}$, M. Trenti$^{5}$, C. M. Carollo$^{4}$, P. G. van Dokkum$^{6}$, M. Franx$^{2}$, M. Stiavelli$^{7}$, V. Gonz\'{a}lez$^{1}$, D. Magee$^{1}$, L. Bradley$^{7}$\\

\noindent

\small \noindent \textit{$^1$Department of Astronomy, University of California Santa Cruz, California 94064, USA}

\small \noindent \textit{$^2$Leiden Observatory, Leiden University, Leiden NL-2333, Netherlands}

\small \noindent \textit{$^3$Carnegie Observatories, Pasadena, California 91101, USA}

\small \noindent \textit{$^4$Institute for Astronomy, ETH Zurich, Zurich CH-8093, Switzerland}

\small \noindent \textit{$^5$University of Colorado, Center for Astrophysics and Space Astronomy, Boulder, Colorado 80303, USA}

\small \noindent \textit{$^6$Department of Astronomy, Yale University, New Haven, Connecticut 06520, USA}

\small \noindent \textit{$^7$Space Telescope Science Institute, Baltimore, Maryland 21218, USA}

\noindent  

\textbf{Searches for very-high-redshift galaxies over the past decade
  have yielded a large sample of more than 6,000 galaxies existing
  just 900-2,000 million years (Myr) after the Big Bang (redshifts
  $6>z>3$; ref. 1). The Hubble Ultra Deep Field (HUDF09) data$^{2,3}$
  have yielded the first reliable detections of $z\approx 8$
  galaxies$^{3-9}$ that, together with reports of a $\gamma$-ray burst
  at $z\approx 8.2$ (refs 10, 11), constitute the earliest objects
  reliably reported to date. Observations of $z\approx 7$-8 galaxies
  suggest substantial star formation at $z>9$-10 (refs 12, 13). Here
  we use the full two-year HUDF09 data to conduct an ultra-deep search
  for $z\approx 10$ galaxies in the heart of the reionization epoch,
  only 500 Myr after the Big Bang. Not only do we find one possible $z
  \approx 10$ galaxy candidate, but we show that, regardless of source
  detections, the star formation rate density is much smaller
  ($\sim$10\%) at this time than it is just $\sim$200 Myr later at
  $z\approx 8$. This demonstrates how rapid galaxy build-up was at
  $z\approx 10$, as galaxies increased in both luminosity density and
  volume density from $z\approx 8$ to $z\approx 10$. The 100-200 Myr
  before $z\approx 10$ is clearly a crucial phase in the assembly of
  the earliest galaxies.}

The detection of galaxies at very high redshift from deep imaging data
depends on the absorption (by intervening neutral hydrogen) of much of
the flux in the spectrum at wavelengths below the wavelength of Lyman
$\alpha$ (121.6 nm). These ‘spectral breaks’ shift to longer
wavelengths for more distant, redshifted galaxies seen at earlier
times. A distinguishing characteristic of $z \approx 10$ galaxies
would be, first, a detection in the $H_{160}$ band, and, second, the
absence of flux in the $J_{125}$ band, and in all other
shorter-wavelength Hubble Space Telescope (HST) Wide Field Camera 3
(WFC3/IR) and Advanced Camera for Surveys (ACS) filters blueward of
the $J_{125}$ band (hence they are called ‘$J_{125}$-dropouts’). The
new, powerful HST WFC3/IR camera is $\sim$40 times more efficient at
finding $z \approx 7$ galaxies$^{2,4-9}$ than the previous near-infrared
NICMOS camera owing to its wider field of view and greater sensitivity
in its $Y_{105}$, $J_{125}$ and $H_{160}$ filters. It provides us with
the capability to explore to $z\approx 10$.

A thorough search of the deep WFC3/IR HUDF09 data set strong limits at
$z \approx 10$, and also resulted in the detection of a candidate
$z\approx 10$ $J_{125}$-dropout galaxy UDFj-39546284 at 5.4$\sigma$ in
our 0.26$''$-diameter selection aperture (Fig. 1). The signal-to-noise
ratio grows to 5.8$\sigma$ in a larger 0.35$''$-diameter aperture.
The candidate is 28.92$\pm$0.18 mag in the WFC3/IR $H_{160}$ band
((1.01$\pm$0.18) $\times$ $10^{31}$ erg s$^{-1}$ cm$^{-2}$ Hz$^{-1}$),
has a likely redshift of $z\approx 10.3$ (Fig. 2), and appears to be
slightly extended. Given the importance of the limits that we set, and
of the candidate $z\approx 10$ galaxy, we perform extensive tests and
simulations. These are described in Supplementary Information sections
4 and 7, while the candidate properties are given in Supplementary
Table 1.

The existence of galaxies at $z> 8.2$ (the $\gamma$-ray burst
redshift$^{10,11}$) is strengthened by three additional sources that
have been detected in recent searches$^{3-9}$, one of which has a
tentative spectroscopic confirmation at $z \approx 8.6$ (ref. 14). The
updated redshift distributions from our simulations show that these
three sources$^3$ are most likely to be at $z \approx 8.7$, 8.5 and
8.6 (Fig. 2). The expectation of finding galaxies at $z\approx 10$,
just $\sim$120 Myr earlier, is enhanced by these strong detections at
$z\approx 8.5$, especially since the $z \approx 7$-8 Spitzer and HST
data suggest that substantial star formation is likely at $z> 9$-10
(refs 12, 13).

The photometric-selection ‘dropout’ approach has been verified through
numerous spectroscopic confirmations at redshifts from $z\approx 2$ to
$z\approx 6$ (refs 15-19), and possibly also now at $z\approx 8.6$
(ref. 14). For our candidate $z\approx 10$ galaxy, however, its single
band ($H_{160}$ band) detection increases the risk of contamination
compared to the $z\approx 7$ and $z\approx 8$ samples, where two (or
more) bands are used to measure the source magnitudes and
colours. Fortunately, we can test the robustness of the single-band
detection process by selecting $z\approx 8$ galaxy candidates using
the $J_{125}$-band data alone. Analogous to the $z\approx 10$
$J_{125}$-dropouts, $z\approx 8$ galaxies are $Y_{105}$-band
dropouts. We compare this single-band selection against the more
robust $z\approx 8$ detections using two bands$^3$ ($J_{125}$ and
$H_{160}$). We are very encouraged that we select the same eight
$z\approx 8$ $Y_{105}$-dropouts with the $J_{125}$-band data alone, as
we do with the normal selection using the $J_{125}$ and $H_{160}$-band
data.  The primary reason for the robustness is the non-detection in
all shorter wavelength filters. The $\chi_{opt} ^2$ test that we have
developed$^9$ largely eliminates contaminating objects.

Our $z\approx 10$ candidate was also checked for any Spitzer IRAC flux
in the 3.6$\mu$m band (see Fig. 1). It is quite isolated and is not
detected to $\sim$27 AB mag (2$\sigma$), further enhancing the case
that this $z\approx 10$ candidate corresponds to a very-high-redshift
galaxy rather than a highly reddened, lower-redshift
contaminant. Contamination from spurious sources is also an important
concern for such faint sources. We verified that the source is present
in a wide variety of subsets of the $H_{160}$-band data (Fig. 1,
Supplementary Fig. 2), suggesting that the candidate is not
spurious. Although these tests make the case for this source being a
$z\approx 10$ galaxy, deeper observations -- involving both imaging
(with, for example, WFC3/IR) and spectroscopy (with the James Webb
Space Telescope) -- will be required to confirm it.

Using the results of these tests and Monte Carlo simulations, we
estimate there is an $\sim$20\% probability that our candidate is a
contaminant or is spurious.  Of that 20\%, 10\% was estimated to be
from photometric scatter. Contamination from spurious sources is
uncertain, and our estimates range from 1\% to 10\% probability; to be
conservative we adopt 10\%. Contamination from lower-redshift red
sources is also possible, but the above single-filter $z\approx 8$
$Y_{105}$-dropout test suggests that the probable contamination is
small, consistent with the totals we estimate from our other tests
($\sim$20\%).

Fortunately, the depth of the data and the thoroughness of our
analysis for contamination allow us to set very strong constraints on
the volume density of $z\approx 10$ galaxies, regardless of the
uncertainties associated with our candidate $z\approx10$ galaxy.  Thus
we evaluate our constraints based on the limit set if no galaxies were
found, and based on the candidate $z\approx 10$ galaxy, whose volume
density is corrected by 20\% to account for the estimated
contamination rate.

Large-scale structure uncertainties are important for small-area
searches. We estimate the field-to-field variance on the present
$z\approx 10$ $J_{125}$-dropout searches in the HUDF09 field to be
39\% (see Supplementary Information)$^{20}$. Even at this level, the
cosmic variance (‘large-scale structure’) is not the dominant source
of uncertainty for our single candidate galaxy.

These $z\approx 10$ results have far-reaching implications for
estimating the role of galaxies in reionization (using the luminosity
density), and for establishing the star formation rate density at very
early times, as $z\approx 10$ is just 480 Myr after the Big Bang and
just a few hundred million years since the first galaxies formed.  The
upper limits and our candidate allow us to do this through
quantitative constraints we place on the $z\sim 10$ luminosity
function.

The extent of the changes at $z\approx 10$ can be demonstrated by
first contrasting what we see at $z\approx 10$ with expectations based
on a no-evolution scenario: that is, the galaxy populations stay
unchanged with time.  We compute the ‘no-evolution’ estimate by using
our ‘galaxy cloning’ software$^{21}$ to artificially redshift the
observed $z\approx 6$ and $z\approx 7$ galaxy population to $z\approx
10$, add them at random positions within our HUDF data, and then
repeat the object selection process just as for the observed $z\approx
10$ galaxy sample.  We estimated that we would find 12$\pm$4 $z\approx
10$ galaxies using our $z\approx 7$ detections as the baseline, and
23$\pm$5 $z\approx 10$ galaxies using our $z\approx 6$ detections as a
baseline. These ‘no-evolution’ estimates are substantially higher than
our (contamination-corrected) estimate of $\sim$0.8 $z\approx 10$
galaxies. For simple Poissonian statistics, our observed number of
$\sim$0.8 galaxies is inconsistent with no-evolution at 4$\sigma$ and
5$\sigma$ confidence, respectively (and sets even stronger limits on
any ‘upturn’ in the star formation rate$^6$). Although striking, this
is not wholly unexpected. Extrapolating the trends seen by us at lower
redshifts$^{22}$ would lead us to expect $3\pm 2$ $z\approx 10$
sources. Thus our results reaffirm that the significant evolution seen
in galaxies at lower redshift continues to $z\approx 10$ (in contrast
with other studies$^6$).

The present search results can also be expressed as constraints on the
luminosity function at $z\approx 10$. The luminosity function
describes the number density of galaxies versus luminosity, and is
important for estimating the ultraviolet flux from galaxies and their
expected role in reionizing the Universe. The high-redshift-galaxy
ultraviolet luminosity function maintains a nearly constant form and
evolves in a largely self-similar manner, with the characteristic
luminosity ($L^*$) increasing smoothly over about 1,300 Myr from
$z\approx 7$ to $z\approx 3$, that is, from $\sim$750 Myr to
$\sim$2,000 Myr.  Assuming the same form for the ultraviolet
luminosity function at z$\sim$10, we find that $L^*$ at $z\approx 10$
is fainter, indicating that the evolution in the bright end of the
ultraviolet luminosity function seen from $z\approx 7$ to $z\approx 4$
(refs 1, 9, 22) continues to $z\approx 10$ (Fig. 3). Definitive
measurements of $L^*$ at $z\approx 10$ will, of course, require deep,
wide-area data to define the luminous end of the $z\approx 10$
luminosity function.

The existence of a steep slope $\alpha$ to the faint end of the
ultraviolet luminosity function found at $z\approx 6$-7 (refs 1, 2, 9)
highlights the importance of low luminosity galaxies in providing the
flux needed to reionize the Universe. It is of great interest to
estimate the ultraviolet luminosity density at $z\approx 7$-10 where
reionization most probably occurred, given its apparent completion at
$z\approx 6$ (ref. 23) and its onset at $z\approx 11$ as deduced from
Wilkinson Microwave Anisotropy Probe (WMAP)$^{24}$ observations.  The
recent results from the HUDF09 data set provide estimates for the
ultraviolet luminosity density at $z \approx 7$ and at $z\approx 8$
(ref. 9). We can now also do so at $z\approx 10$.  We compute the
luminosity density implied by our sample by assuming a faint-end slope
of $-$1.7 (the same slope as found for the $z\approx 2$-7 luminosity
functions) and extending the integration down to a very plausible
limit of $-$12 AB mag. We find that the ultraviolet flux that is
available from galaxies at $z\approx 10$ is only
$\sim$12$_{-10}^{+26}$\% of what is needed for galaxies to be the
reionizing source, with typical assumptions of an escape fraction of
$\sim$0.4, a clumping factor of $\sim$3 and a Salpeter initial mass
function (see, for example, ref. 9). This result is tantalizing,
suggesting that galaxies are contributing to reionization, but the
enigma remains: where are most of the needed ultraviolet photons
coming from?  Observations to significantly fainter levels will be
central to characterizing the role of galaxies in reionization.

The star formation rate (SFR) density is derived from the luminosity
density (see Fig. 4).  The SFR density increases systematically and
monotonically at early times from $z\approx 10$ (500 Myr) to $z\approx
4$ (1,600 Myr), peaking at $z\approx 2$-3 (at $\sim$2,500 Myr), before
decreasing at $z< 2$ (Fig. 4). This suggests that the luminosity
function and star formation rate density evolution found at lower
redshifts$^{1,25}$ continues to $z\approx 10$ when the universe was
just 480 Myr old. The limits established here even suggest that the
trends in star formation rate density established at lower redshifts
could be steepening.

This is clearly an era when galaxies were evolving very rapidly.  The
star formation rate density increased by a factor of $\sim$10 in less
than 200 Myr, from $z\approx 10$ to $z\approx 8$. This dramatic change
in such a short period of time suggests that the first phases of
galaxy formation and their build-up could be unveiled by observations
that penetrate just 200 Myr earlier, to redshifts $z\approx
15$. However, only when the James Webb Space Telescope is launched
will these first phases of galaxy build-up between $z\approx 15$ and
$z\approx 10$ be revealed.

\noindent \textbf{Received 21 December 2009}\\

  \noindent 1.  Bouwens, R.J., Illingworth, G.D., Franx, M., \& Ford, H.  UV Luminosity Functions at $z\sim4$, 5, and 6 from the Hubble Ultra Deep Field and Other Deep Hubble Space Telescope ACS Fields: Evolution and Star Formation History. \textit{Astrophys. J.} \textbf{670}, 928-958 (2007).

  \noindent 2.  Oesch, P.A., \textit{et al. } $z\sim7$ Galaxies in the
  HUDF: First Epoch WFC3/IR Results. \textit{Astrophys. J. Lett.}
  \textbf{709}, L16-L20 (2010).

  \noindent 3.  Bouwens, R.J., \textit{et al.} Discovery of $z\sim8$ Galaxies in the HUDF from ultra-deep WFC3/IR Observations. \textit{Astrophys. J. Lett.} \textbf{709}, L133-L137 (2010).

  \noindent 4.  McLure, R., \textit{et al.} Galaxies at $z\sim6-9$ from the WFC3/IR imaging of the HUDF. \textit{Month. Not. R. Astron. Soc.} \textbf{403}, 960-983 (2010).

  \noindent 5.  Bunker, A., \textit{et al.} The Contribution of High Redshift Galaxies to Cosmic Reionization: New Results from Deep WFC3 Imaging of the Hubble Ultra Deep Field. \textit{Month. Not. R. Astron. Soc.} \textbf{409}, 855-866 (2010).

  \noindent 6.  Yan, H., \textit{et al.  }Galaxy Formation In The Reionization Epoch As Hinted By Wide Field Camera 3 Observations Of The Hubble Ultra Deep Field. \textit{Res. Astron. Astrophys.} \textbf{10}, 867-904 (2010).

  \noindent 7.  Finkelstein, S. \textit{et al.  }On the Stellar Populations and Evolution of Star-Forming Galaxies at $6.3<z<8.6$. \textit{Astrophys. J.} \textbf{719}, 1250-1273 (2010).

  \noindent 8.  Robertson, N., Ellis, R. S., Dunlop, R. S., McLure, R. J. \& Stark, D. P. Early star-forming galaxies and the reionization of the Universe. \textit{Nature} \textbf{468}, 49-55 (2010).

  \noindent 9.  Bouwens, R. J. \textit{et al.}  UV luminosity
  functions from 113 z$\sim$7 and z$\sim$8 Lyman-break galaxies in the
  ultra-deep HUDF09 and wide-area ERS WFC3/IR observations.
  2010. \textit{Astrophys. J.} (submitted); preprint at
  $<$http://arxiv.org/abs/1006.4360$>$ (2010).

  \noindent 10.  Tanvir, N., \textit{et al.}  A $\gamma$-ray burst at a redshift of $z\sim8$. \textit{Nature} \textbf{461}, 1254-1257

  \noindent 11.  Salvaterra, R., \textit{et al.}  GRB090423 at a redshift of $z\sim8.1$. \textit{Nature} \textbf{461}, 1258-1260

  \noindent 12.  Labbe, I., \textit{et al.}  Ultradeep IRAC Observations of sub-$L^*$ $z\sim7$ and $z\sim8$ Galaxies in the HUDF: the Contribution of Low-Luminosity Galaxies to the Stellar Mass Density and Reionization. \textit{Astrophys. J. Lett.} \textbf{708}, L26-L31 (2010).

  \noindent 13.  Gonzalez, V. \textit{et al.}  Stellar Mass Density and Specific Star Formation Rates of the Universe at $z\sim7$, \textit{Astrophys. J.} \textbf{713}, 115-130 (2010).

\noindent 14.  Lehnert, M. \textit{et al.\ } Spectroscopic confirmation
  of a galaxy at $z\sim8.6$.  \textit{Nature} \textbf{467}, 940-942 (2010).

\noindent 15.  Steidel,  C.C., Giavalisco, M., Pettini, M., Dickinson, M., \& Adelberger, K.L.  Spectroscopic Confirmation of a Population of Normal Star-forming Galaxies at Redshifts $Z>3$. \textit{Astrophys. J. Lett.} \textbf{462}, 17-20 (1996).

  \noindent 16.  Vanzella, E., \textit{et al.  }Spectroscopic Observations of Lyman Break Galaxies at Redshifts $\sim$4, 5, and 6 in the Goods-South Field.  \textit{Astrophys. J.} \textbf{695}, 1163-1182 (2009).

  \noindent 17.  Popesso, P. \textit{et al.} The great observatories
  origins deep survey. VLT/VIMOS spectroscopy in the GOODS-south
  field. \textit{Astron. Astrophys.} \textbf{494}, 443-460 (2009).

  \noindent 18.  Steidel, C.C. \textit{et al.}  Lyman Break Galaxies at
  Redshift $z\sim3$: Survey Description and Full Data Set.
  \textit{Astrophys. J.} \textbf{592}, 728-754 (2003).

  \noindent 19.  Reddy, N. \textit{et al.}  A Spectroscopic Survey of
  Redshift $1.4<\sim z<\sim3.0$ Galaxies in the GOODS-North Field:
  Survey Description, Catalogs, and Properties. \textit{Astrophys. J.}
  \textbf{653}, 1004-1026 (2006).

  \noindent 20.  Trenti, M. \& Stiavelli, M.  Cosmic Variance and its Effect on the Luminosity Function Determinations in Deep High-z Surveys. \textit{Astrophys. J.} \textbf{676}, 767-780 (2008).

  \noindent 21.  Bouwens, R.J., Broadhurst, T.J., Silk, J.  Cloning Hubble Deep Fields. I. A Model-independent Measurement of Galaxy Evolution. \textit{Astrophys. J.} \textbf{506}, 557-578 (1998).

  \noindent 22.  Bouwens, R.J., Illingworth, G.D., Franx, M., \& Ford, H. z $\sim$ 7-10 Galaxies in the HUDF and GOODS Fields: UV Luminosity Functions. \textit{Astrophys. J.} \textbf{686}, 230-250 (2008).

  \noindent 23.  Fan, X., \textit{et al.}  Evolution of the Ionizing Background and the Epoch of Reionization from the Spectra of $z\sim6$ Quasars. \textit{Astronom. J.} \textbf{123}, 1247-1257 (2002).

  \noindent 24.  Komatsu, E., \textit{et al.} Seven-Year Wilkinson
  Microwave Anisotropy Probe Observations: Cosmological
  Interpretation. \textit{Astrophys. J.} (in press); preprint at
  $<$http://arXiv.org/abs/1001.4538$>$ (2010).

  \noindent 25.  Bouwens, R., \textit{et al.}  UV-Continuum Slope and Dust Obscuration from $z\sim6$ to $z\sim2$: The Star Formation Rate Density at High Redshift. \textit{Astrophys. J.} \textbf{705}, 936-961 (2009).

  \noindent 26.  Beckwith, S.W., \textit{et al.}  The Hubble Ultra Deep Field. \textit{Astrophys. J.} \textbf{132}, 1729-1755 (2006).

  \noindent 27.  Reddy, N., \& Steidel, C.C.   A Steep Faint-End Slope of the UV Luminosity Function at z $\sim$ 2-3: Implications for the Global Stellar Mass Density and Star Formation in Low-Mass Halos. \textit{Astrophys. J.} \textbf{692}, 778-803 (2009).

  \noindent 28.  Yoshida, M., et al.  Luminosity Functions of Lyman Break Galaxies at $z\sim4$ and $z\sim5$ in the Subaru Deep Field. \textit{Astrophys. J.} \textbf{653}, 988-1003 (2006).

  \noindent 29.  McLure, R., Cirasuolo, M., Dunlop, J.S., Foucaud, S., Almaini, O.  The luminosity function, halo masses, and stellar masses of luminous Lyman-break galaxies at $5<z<6$. \textit{Month. Not. R. Astron. Soc.} \textbf{395}, 2196-2209 (2009).

  \noindent 30.  Schiminovich, D., \textit{et al.}  The GALEX-VVDS Measurement of the Evolution of the Far-Ultraviolet Luminosity Density and the Cosmic Star Formation Rate. \textit{Astrophys. J. Lett} \textbf{619}, 47-50 (2005).

\noindent \textbf{Supplementary Information} is linked to the online version of the paper at www.nature.com/nature.

\noindent \textbf{Acknowledgements} We are grateful to all those at NASA, STScI and throughout the community who have worked to make the Hubble Space Telescope the observatory that it is today, and we acknowledge the importance of the servicing missions and those who organised them. We acknowledge our program coordinator W. Januszewski for his care in helping to set up our program and observing configuration. We acknowledge support from NASA and the Swiss National Science Foundation.\\

\noindent \textbf{Author Contributions} R.J.B. carried out the most of the data analysis and calculations for this paper, and wrote most of the Supplementary Information; G.D.I. wrote most of the text in the Letter and iterated on the initial science results and content with R.J.B.; I.L., P.A.O., M.T., C.M.C., P.G.v.D., M.F., M.S. and L.B. provided significant feedback on the science content and on the drafts; I.L. and V.G. were involved with processing the Spitzer IRAC data; P.A.O. contributed to the data analysis; M.T. made the cosmic variance estimates; and D.M. was involved in data processing and pipeline generation for the WFC3/IR data.

\noindent \textbf{Author Information} Reprints and permissions information is available at www.nature.com/reprints. The authors declare no competing financial interests. Readers are welcome to comment on the online version of this article at www.nature.com/nature. Correspondence and requests for materials should be addressed to R.J.B. (bouwens@ucolick.org).

  \noindent \includegraphics[width=6in]{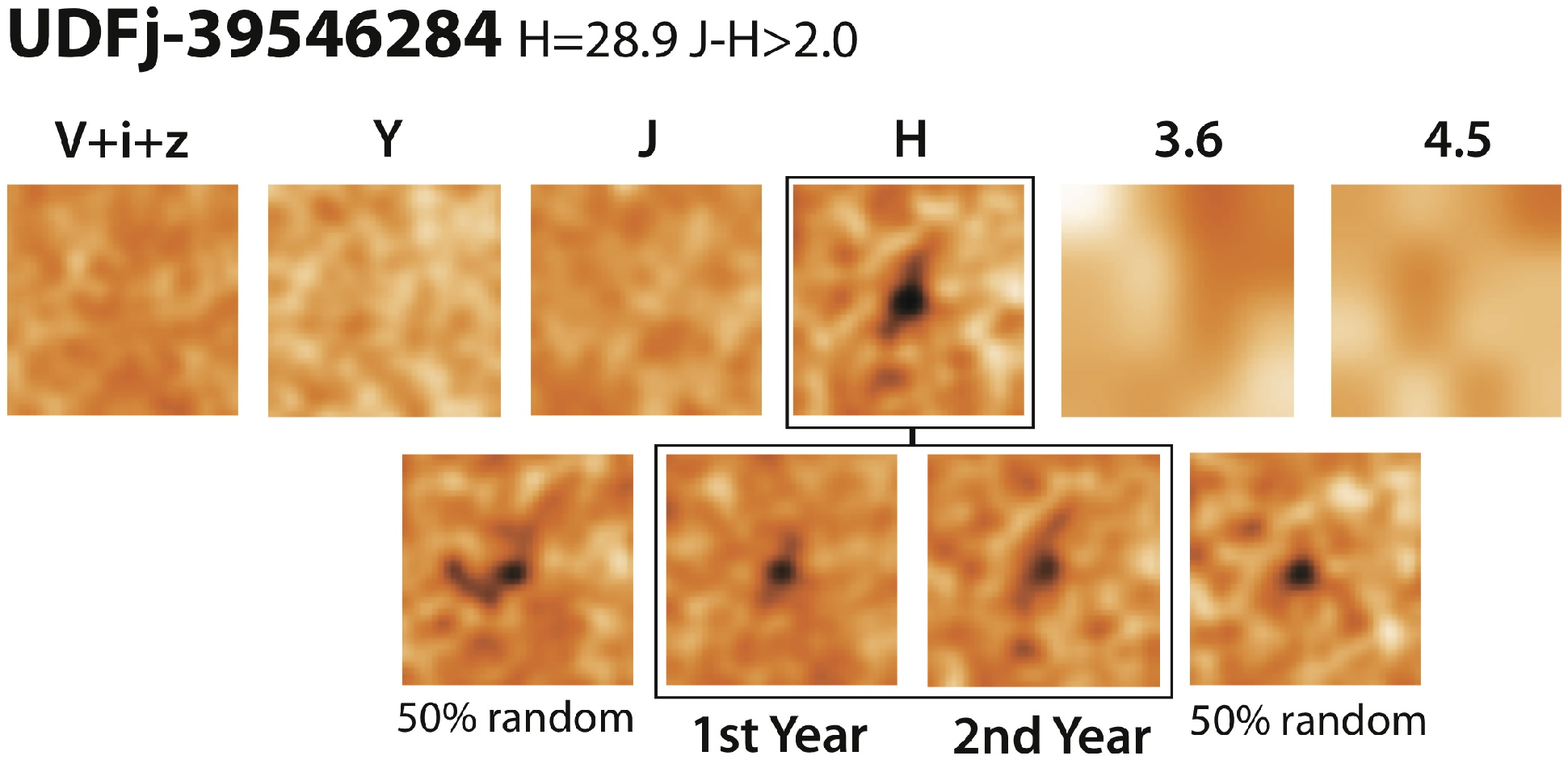}

\noindent Figure 1.  Optical and near-infrared images of the candidate
$z\approx 10$ galaxy, UDFj-39546284, from the HUDF. Top row: the
leftmost panel shows the HUDF ACS ($V_{606}i_{775}z_{850}$)
data$^{26}$; the next three panels show the similarly deep HUDF09 (HST
GO 11563), near-infrared WFC3/IR ($Y_{105}$, $J_{125}$, $H_{160}$)
data (reaching to 5$\sigma$ depths of $\sim$29.8 AB mag)$^{2,3,9}$;
and the last two panels show the longer wavelength Spitzer IRAC 3.6
and 4.5 $\mu$m observations. Bottom row: the two middle panels show
images of our $z\approx 10$ candidate in the first and second year of
$H_{160}$-band observations (each representing $\sim$50\% of the
total); the two outer panels show two random 50\% subsets of the data
(see also Supplementary Fig. 2). Each cutout is 2.4$''$ $\times$
2.4$''$ on a side, and is orientated with north at the top. For our
selection criteria, we require our $z\approx 10$ candidates to be
detected at 5$\sigma$ in the $H_{160}$ band, to have $J_{125}-H_{160}$
colours redder than 1.2 AB mag, and to be undetected ($<$2$\sigma$) in
all imaging observations blueward of the $J_{125}$ band. Also,
candidates must not be detected at $>$1.5$\sigma$ in more than one
band blueward of the $J_{125}$ band, and have $\chi^2 < 2.5$ in the
extremely-deep image obtained by combining the $B_{435}$, $V_{606}$,
$i_{775}$, $z_{850}$ and $Y_{105}$ imaging data. All of these
requirements place very strong limits on the optical flux from our
$z\approx 10$ candidates and provide strong discrimination against
contamination by low-redshift sources (see ref. 9, Appendix C). The
candidate is significant at $>$3$\sigma$ in each year of observations
and therefore not likely to be spurious.  It is detected at
5.4$\sigma$ in the $H_{160}$ band, which is much more significant than
the next possible candidates (seen at 4.0$\sigma$ and 4.9$\sigma$). In
addition, our $z\approx 10$ candidate is not detected in the IRAC
data, as expected given the IRAC flux limits. The position and other
properties of this candidate are given in Supplementary Table 1.

  \noindent \includegraphics[width=6in]{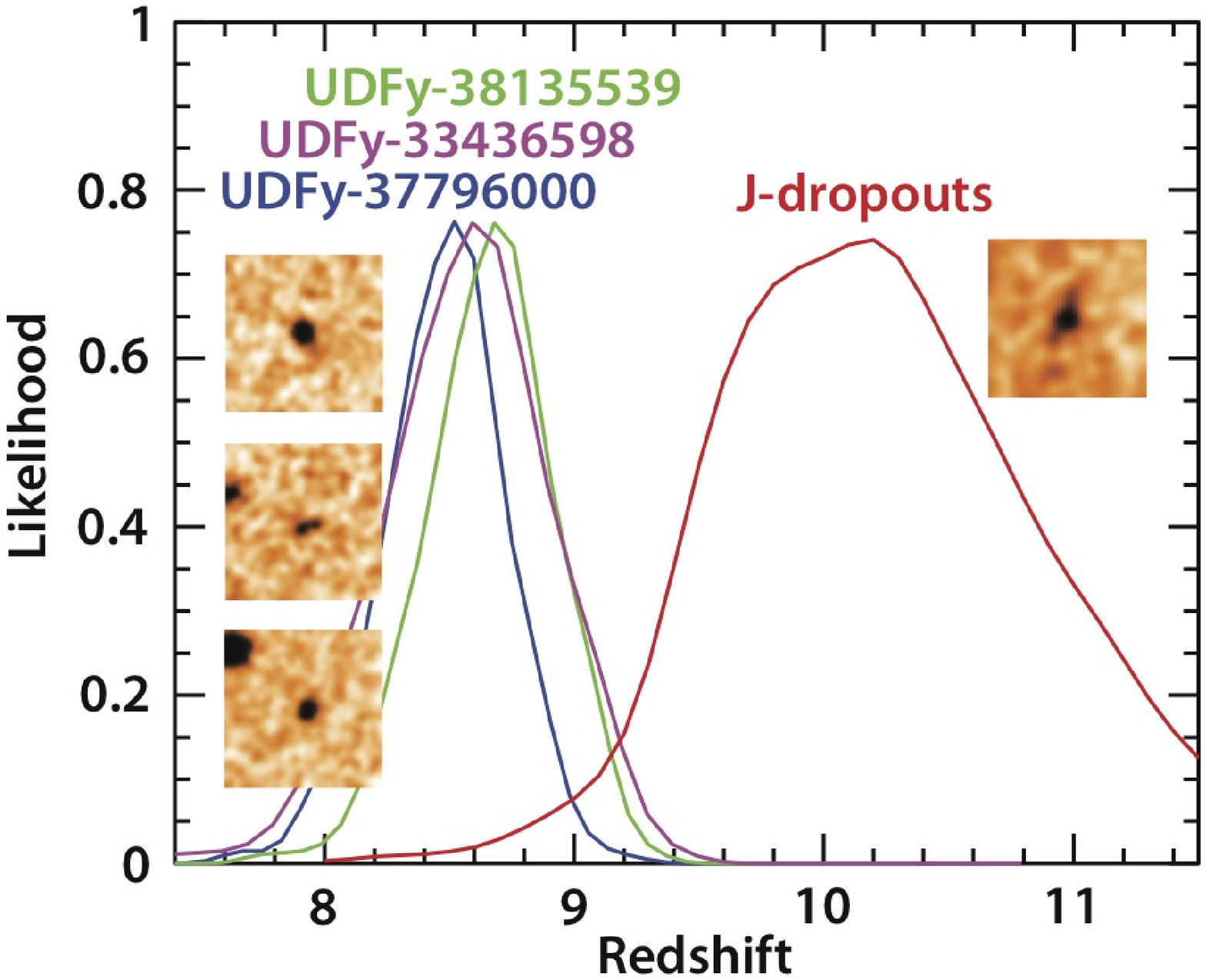}

\noindent Figure 2. Predicted redshift distributions for our $z\approx
8.5$ and $z\approx 10$ galaxy candidates. The red line gives the
redshift distribution for our $z\approx 10$ $J_{125}$-dropout
candidate, while the blue, green and magenta lines give the redshift
distributions for our $z\approx 8.5$ $Y_{105}$-dropout candidates. The
$H_{160}$-band source images that correspond to the redshift
distributions are shown for the $z\approx 8.7$, 8.6 and $z\approx 8.5$
sources (labels arranged in same order as images), and for our
$z\approx 10.3$ candidate. Each source image is 2.4$''$ $\times$
2.4$''$, with north at the top. The selection criteria for the z
$\approx$ 8-9 sources have been published elsewhere$^{3,9}$ (see also
Supplementary Information section 5).  However, the detailed redshift
distributions are shown here for the first time. The redshift
distributions were derived by adding artificial sources to the HUDF09
WFC3/IR data, and reselecting them in the same way as the actual
galaxy candidates (see Supplementary Information sections 4, 5 and
9). The mean redshifts of these distributions are 8.7, 8.6, 8.5 and
10.3. The $z\approx 8.7$ source has a tentative spectroscopic
confirmation at $z\approx 8.6$ (ref. 14). For these simulations, the
ultraviolet luminosity function was used.  The luminosity function
describes the number density of galaxies versus luminosity and is
usually parameterized as $\phi^* e^{-L/L^*}(L/L^*)^{\alpha}$, where
$\phi^*$ is the normalization, $L^*$ is the characteristic luminosity,
and $\alpha$ is the faint end slope (see Fig. 3). The luminosity
function was assumed to have an $M_{UV} ^{*}$ of $-$19.5 and $-$18.8
at $z\approx 8$ and $z\approx 10$, respectively (based on predictions
from our z $\approx$ 4-7 fitting relation$^{9,22}$), while $\alpha$
was taken to be $-$1.74.

  \noindent  

  \noindent \includegraphics[width=6in]{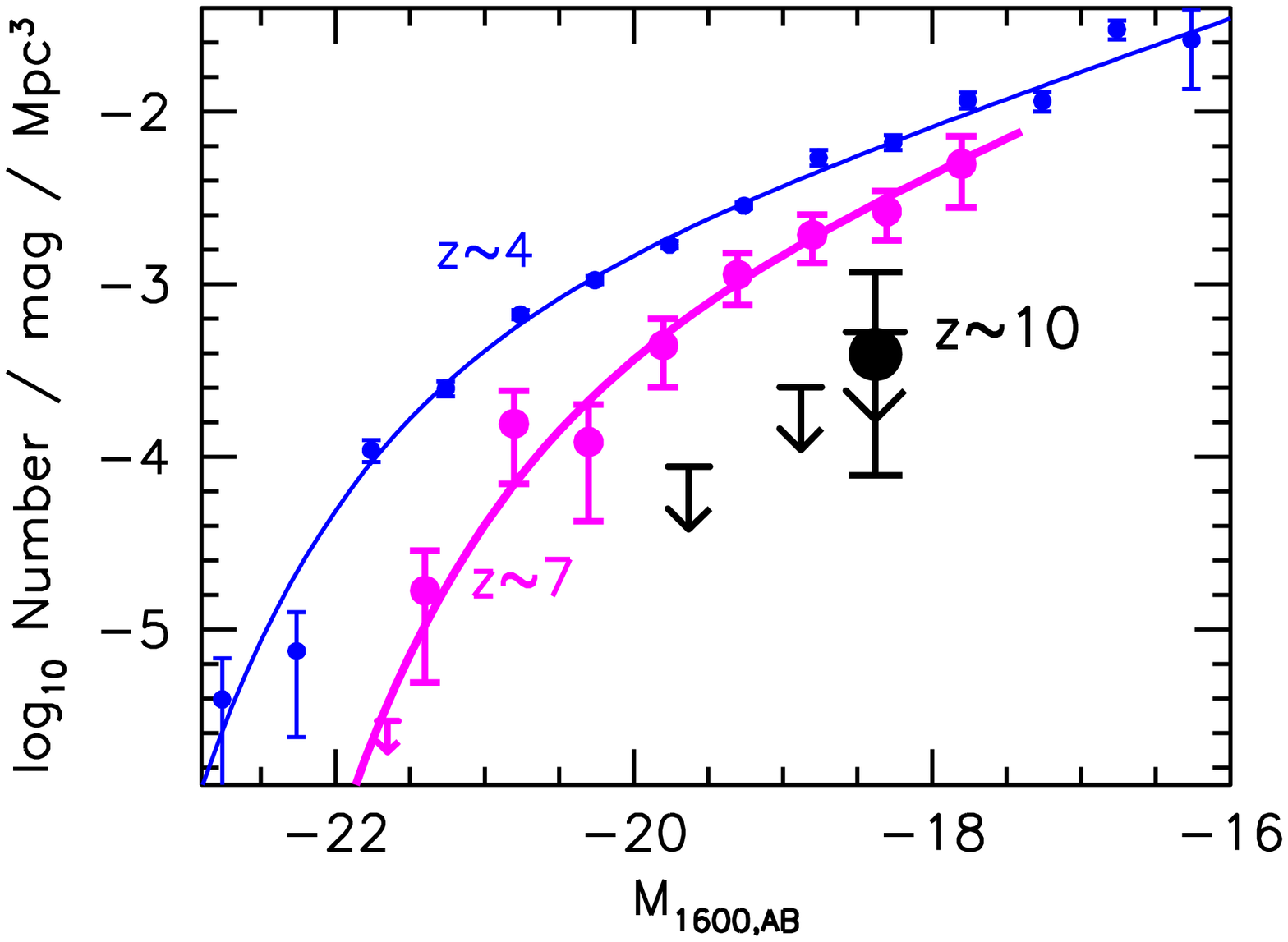}

\noindent Figure 3.  Ultraviolet luminosity functions at $z\approx 4$,
$z\approx 7$ and constraints for $z\approx 10$. The present
constraints on the stepwise ultraviolet luminosity function at
$z\approx 10$ (black points and upper limits) are new and are derived
from the $J_{125}$-dropout candidate galaxies over our ultra-deep HUDF
WFC3/IR field. These luminosity functions are a function of the
absolute magnitude (that is, luminosity) of galaxies ($M_{1600,AB}$)
in the rest-frame far-ultraviolet. All error bars are 1$\sigma$. The
stepwise luminosity function at $z\approx 10$ is also presented as a
1s upper limit, given the uncertainty of our $z\approx 10$
candidate. The lowest luminosity point has been corrected for
incompleteness. The ultraviolet luminosity functions$^{1,9}$ at
$z\approx 4$ (blue) and at $z\approx 7$ (magenta) are shown for
comparison. The luminosity functions fitted here are of the form
$\phi^* e^{-L/L^*} (L/L^*)^{\alpha}$ (see Fig. 2 legend). This
analytic representation has recently been shown$^{9}$ to fit well at
$z\sim 7$ and later times. The present search results also allow us to
estimate the value of $L^*$ at $z\approx 10$ -- assuming that the
luminosity function at $z\approx 10$ has the same values of $\phi^*$
and $\alpha\approx -1.7$ as have been found to describe ultraviolet
luminosity function results from $z\approx 7$ to $z\approx 4$ (refs 1,
2, 9, 27-29). Doing so allows us to constrain the evolution in the
luminosity function out to $z\approx 10$, nearly 500 Myr earlier than
at $z\approx 6$ (and so halving the time difference between the first
galaxies at $z\approx 15$-20 and those seen at $z\approx 6$). We find
$L^*$ at $z\approx 10$ to be $-$18.3$\pm$0.5 AB mag, or $L^* > -18.3$
in the limit of no detected sources -- although obviously very
uncertain, this is consistent with the evolution in the bright end of
the ultraviolet luminosity function seen from $z\approx 7$ to
$z\approx 4$ continuing to $z\approx 10$.

  \noindent \includegraphics[width=6in]{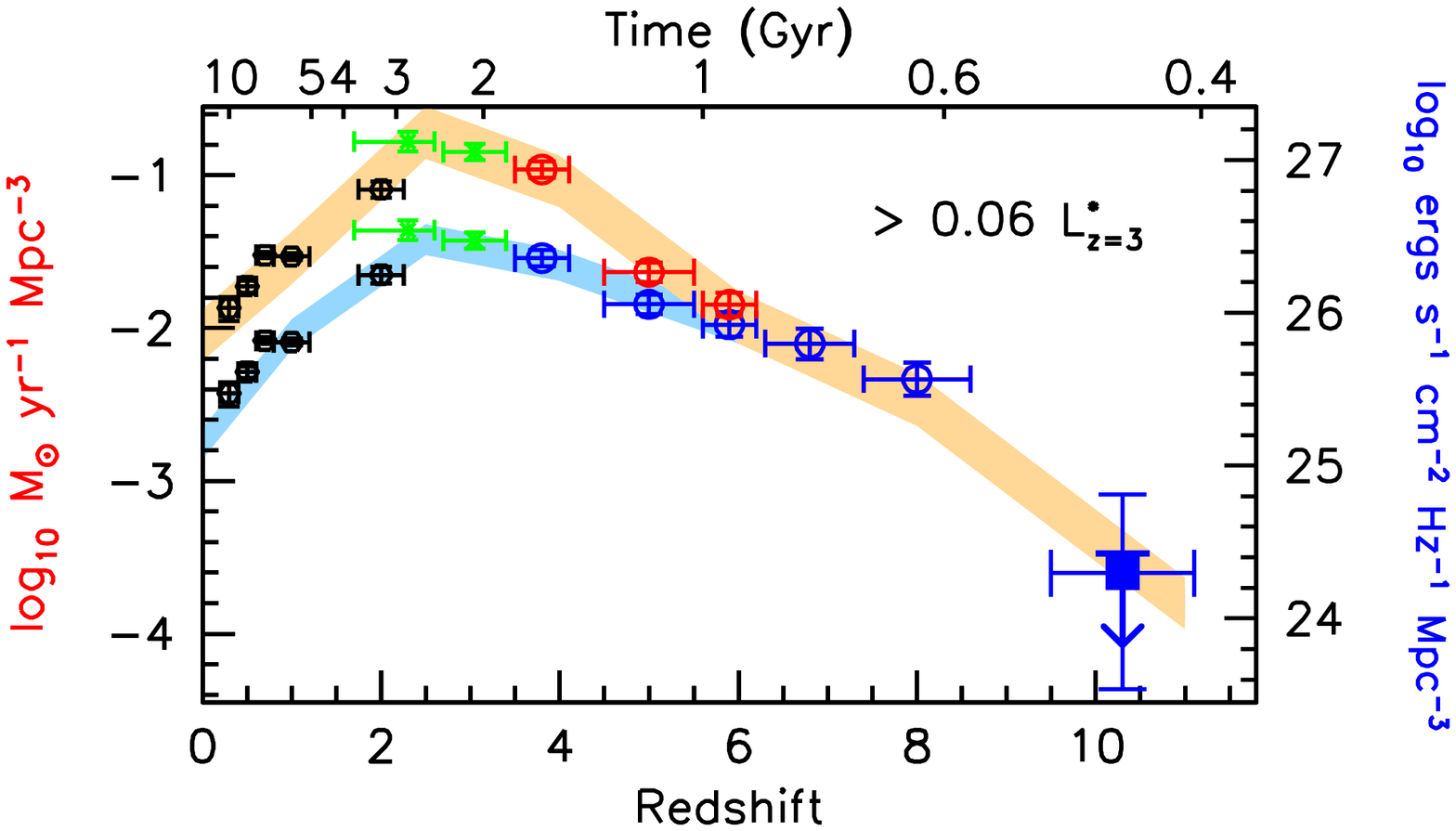}

\noindent Figure 4.  The luminosity density and star formation rate
density in the Universe over 13.2 Gyr. The rest-frame continuum
ultraviolet luminosity density (right axis, blue points) at $z\approx
10$, and the star formation rate density (left axis, red points)
derived from the extinction-corrected luminosity density$^{1,25}$, are
integrated down to the approximate magnitude limit $M_{AB} \approx
-18$ (0.06 L$^{*}$) of our $z\approx 10$ $J_{125}$-dropout search.
The conversion from ultraviolet luminosity to star formation rate
assumes a Salpeter initial mass function.  The upper horizontal axis
gives the time after the Big Bang and the lower axis the redshift. As
before, we assume that the ultraviolet luminosity function has the
same faint-end slope (and normalization) as at $z\approx 6$ and
$z\approx 7$. The star formation rate density ($1.9_{-1.5} ^{+4.4}$
$\times$ $10^{24}$ $M_{\odot}$ yr$^{-1}$ Mpc$^{-3}$) from the
contamination-corrected sample is shown at $z\approx 10$ from the
current $J_{125}$-dropout search, as is the 1$\sigma$ upper limit
($<$3 $\times$ $10^{24}$ $M_{\odot}$ yr$^{-1}$ Mpc$^{-3}$) if we
assume no $z\approx 10$ sources are detected. All error bars are
1$\sigma$.  Also included here are the recent star formation rate
determinations at $z\approx 7$ and $z\approx 8$ from our HUDF09
WFC3/IR $z_{850}$-dropout and $Y_{105}$-dropout searches9, and from
the literature at $z < 4$ (green and black points: refs 27, 30) and at
$z \approx 4$-6 (ref. 1). The dust corrections at $z\approx 4$ are
based on the estimated ultraviolet-continuum slopes $\beta$, and are
already negligible by $z\approx 7$ (refs 2, 3, 25). There is no
evidence for any substantial change in the star formation rate density
trends established at lower redshift.

\clearpage

\begin{center}
\large \textbf{Supplementary Information for Nature Letter\\ Searches and
  limits for z$\sim$10 galaxies in the HST HUDF09 Data\\ R.J. Bouwens,
  G.D. Illingworth and the HUDF09 Team}
\end{center}

\section{General Outline}

Here we describe our organizational plan for the Supplementary
Information to our paper on searches for $z\sim10$ galaxies in the
ultra-deep HUDF09 WFC3/IR data.  In the main section (\S1-\S11), we
present an analysis of the $z\sim10$ and $z\gtrsim8.2$ searches we
performed in the full two-year data set, as well as any conclusions
derived on the basis of those searches.  Included in the main section
is a description of the observational data (\S2), the procedure by
which we construct source catalogs (\S3), our $z\sim10$ and
$z\gtrsim8.2$ searches (\S4, \S5), $z\sim10$ searches in wider-area
observations (\S6), an assessment of possible contamination (\S7),
comparison of the selection from the full two-year data set to the
first-year selection (\S8), the expected number of high-redshift
candidates in our selections (\S9), the implications for high-redshift
LFs (\S10), and implications for the SFR density at $z\sim10$ (\S11).

In the Appendix, we will provide an assessment of the search we
performed for $z\sim10$ galaxies in the first-year HUDF data, where we
reported three possible $z\sim 10$ candidates (arXiv:0912.4263).
These sources have now been checked in the full two-year dataset, and
found to have S/N that is too low to be credible candidates.

\section{Observational Data}

We make use of the ultra-deep two-year 4.7 arcmin$^2$ WFC3/IR
observations over the HUDF from the HUDF09 program (GO 11563: PI
Illingworth).  We also make use of the wide-area 39.2 arcmin$^2$
WFC3/IR observations over the GOODS field from the Early Release
Science (ERS) program (GO 11359: PI O'Connell) to better constrain the
bright end of the $z\sim10$ luminosity function.  The ultra-deep
WFC3/IR observations over the HUDF include 24 orbits, 34 orbits, and
53 orbits of data in the F105W, F125W, and F160W bands, respectively.

Our reductions of the ultra-deep WFC3/IR observations over the HUDF
follow a very similar procedure to that used in our previous
reductions of ultra-deep NICMOS observations$^{22}$ or other deep
near-IR data.  The process begins with individual \textit{flt} files
requested from the archive.  These frames had already been subject to
bias correction, dark subtraction, flat-field correction, and cosmic
ray rejection, so minimal additional calibrations were required.  The
\textit{flt} files obtained from the archive should be reasonably well
calibrated, given that the on-the-fly pipeline processing by STScI
already takes advantage of darks and flat fields constructed from
on-orbit data (the \textit{flt} frames available shortly after launch
were based upon test data taken while still on the ground).

To improve the pixel-by-pixel S/N (and correct for any imperfections
in the flats or darks), we median stacked all of the observations in
the F105W, F125W, and F160W bands to create supermedian frames --
which were subsequently subtracted from the individual images.
Compact, bright point-like sources are then identified in each WFC3/IR
image and in the ACS HUDF data.  These source lists were used to align
the WFC3/IR images with the ACS HUDF data$^{26}$ using our
\textit{superalign} software.$^{31}$ A description of our reductions
of the ultra-deep first-epoch WFC3/IR observations over the HUDF is
already included in Oesch et al.$^2$ and Bouwens et al.$^3$

The WFC3/IR observations were then drizzled onto the HUDF frame using
the \textit{multidrizzle} software$^{32}$ while clipping 4$\sigma$
outliers.  As a result of this clipping procedure, small artifacts in
individual exposures -- such as hot pixels -- will not significantly
affect our reductions.  We attempted to mask out pixels on the WFC3/IR
camera that were affected by source persistence.  This masking was
performed by remapping our initial reductions of the data (or more
precisely blotting median stacks of the data) back to the original
frames, subtracting them from the original frames, coadding these
subtracted frames for all exposures within a visit, smoothing, and
then flagging all pixels above a $3\sigma$ threshold.  This together
with our dither strategy and CR clipping procedure ensures that our
final reductions are not significantly affected by source persistence.

To further optimize the pixel-by-pixel S/N in our reductions (ensuring
they were of the highest quality), we repeated our median stacks of
the exposures in each band -- but now masking out the sources apparent
in our final stacks (rather than just those apparent in the individual
exposures).  We then subtracted these supermedian images off of each
of the individual exposures and drizzled the data together to generate
our final reductions.  The WFC3/IR $Y_{105}$, $J_{125}$, and $H_{160}$
data over the HUDF09 reach to $5\sigma$ depths of 29.6, 29.9, and
29.8, respectively, in $0.35''$-diameter apertures (no correction is
made to these magnitudes to account for the light in sources outside
these apertures).  These depths are estimated from measurements of the
RMS fluctuations in the data itself using apertures of the same radii.
These apertures are placed in blank regions of the data to avoid
obvious sources.  As a result, the noise is determined directly from
the data and so any correlations in the noise are implicitly included
in the estimated depths.  For our selection, we are using smaller
$0.26''$-diameter apertures than for our estimated depths.  This
ensures that our S/N measurements for individual sources are less
sensitive to uncertainties in the estimated sky levels.

\begin{figure}
\epsscale{0.85}
\plotone{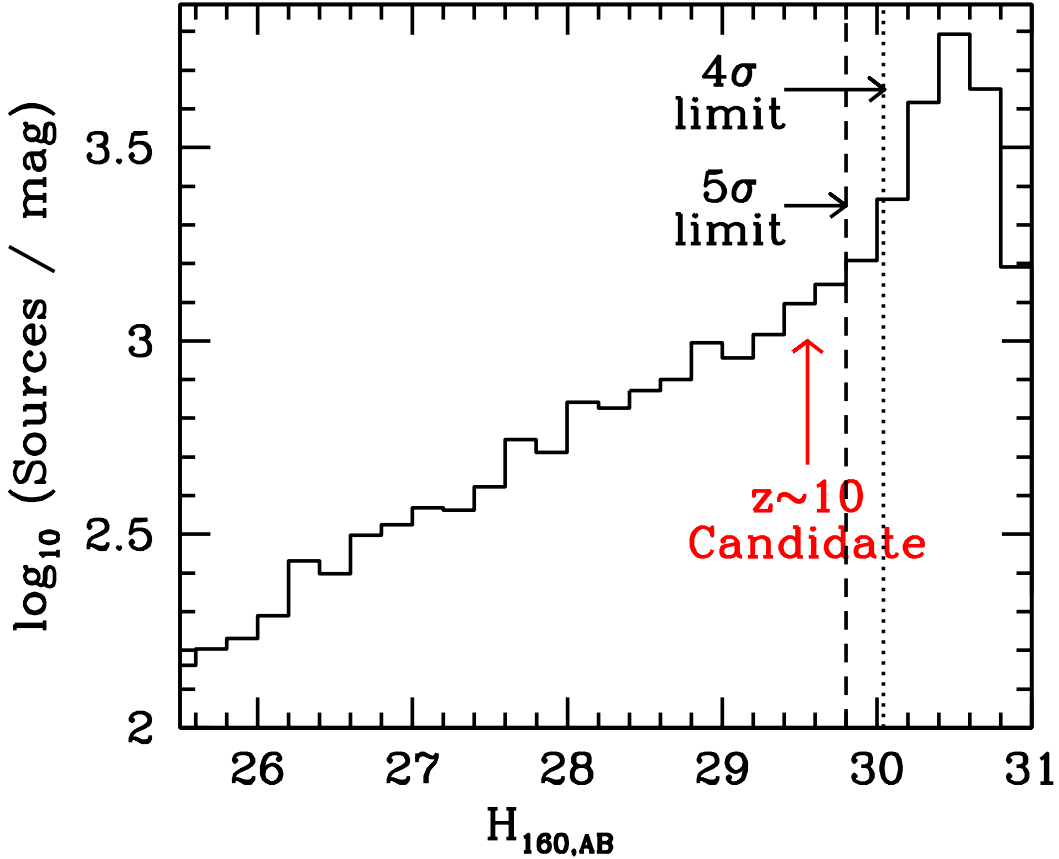}
\caption{Number of sources per unit magnitude as a function of the
  $H_{160}$-band magnitude (measured in $0.35''$-diameter) in our
  source catalogs.  No correction for flux outside this aperture and
  on the wings of the PSF have been made, to compare with our
  $5\sigma$ limits (29.8 AB mag) quoted in the same manner and shown
  here with a dotted vertical line along with the $4\sigma$ limit.
  Our sole $z\sim10$ $J_{125}$-dropout candidate is a
  $\gtrsim$5.4$\sigma$ source ($0.26''$-diameter aperture: but is
  $\sim$5.8$\sigma$ in the aperture used for the magnitudes plotted in
  the figure) and is $\gtrsim$1.0 mag brighter than the faintest
  sources in our catalogs (which are only $\sim$3$\sigma$, and so
  excluded in our final selection).\label{fig:counts}}
\end{figure}

For the ACS observations over the HUDF, we simply make use of the v1.0
ACS reductions.$^{26}$ The ACS/WFC $B_{435}$, $V_{606}$, $i_{775}$,
and $z_{850}$-band optical data reach to 29.7, 30.1, 29.9, and 29.4 AB
mag, respectively (5$\sigma$: 0.35$''$-diameter apertures).  The FWHMs
of the PSFs are $\sim0.10''$ and $\sim0.16''$ in the ACS $BViz$ and
WFC3/IR $YJH$ observations, respectively.

Our reductions of the WFC3/IR observations over the ERS fields were
performed in a very similar manner to our HUDF09 reductions.  Those
observations reach to 27.9, 28.4, and 28.1 AB mag in the $Y_{098}$,
$J_{125}$, and $H_{160}$ bands, respectively (5$\sigma$:
$0.35''$-diameter aperture).$^9$ The ACS/WFC $B_{435}$, $V_{606}$,
$i_{775}$, and $z_{850}$-band optical data are available from the ACS
GOODS program$^{33}$ and various SNe search and follow-up
programs.$^{34}$ Our reductions of these data$^{1,35}$ are similar to
the GOODS v2.0 reductions, but reach $\sim$0.1-0.3 mag deeper in the
$z_{850}$ band (due to the inclusion of the SNe follow-up
data$^{33}$).  These reductions reach to 28.2, 28.5, 28.0, and 28.0 AB
mag in the $B_{435}$, $V_{606}$, $i_{775}$, and $z_{850}$ bands,
respectively (5$\sigma$: 0.35$''$-diameter apertures).

\section{Source Catalogues}

Our procedure for doing object detection and photometry is essentially
identical to that performed in many previous analyses by our
team.$^{1,22,25,36}$ We provide a brief summary here.  The SExtractor
software$^{37}$ is used for object detection and photometry.  Object
detection is performed using the $H_{160}$-band image.  The ACS
optical images are smoothed to match the WFC3/IR images, and colors
are measured in small scalable apertures (using a Kron$^{38}$ factor
of 1.2).  (A scalable aperture is the area in a two-dimensional
ellipse where the major axis, minor axis, and orientation angle are
set equal to some multiple of the first moments of light in an
object.)  Fluxes in these small scalable apertures are corrected to
total magnitudes, considering the flux in a larger scalable aperture
(Kron factor of 2.5).  This latter correction is made using the
$H_{160}$-band image.  Finally, a correction is made for light outside
this large scalable aperture using the encircled energies tabulated
for point sources in the WFC3/IR F160W band
(http://www.stsci.edu/hst/wfc3/documents/handbooks/currentIHB/c07\_ir7.html).

\section{$J_{125}$-dropout Selection}

We search for high redshift galaxies at $z\sim10$, utilizing
$J_{125}$-dropout selection criteria.  These selection criteria should
permit us to identify candidate $z$$\sim$10 star-forming galaxies in
the ultra-deep WFC3/IR data, should they be brighter than 29.5 AB mag
($>$0.06 $L_{z=3}^{*}$: where this limit implicitly includes a fairly
minimal correction for light on the PSF wings).
Figure~S\ref{fig:counts} shows the number of sources we find in the
WFC3/IR HUDF09 observations over the HUDF versus $H_{160}$-band
magnitude.

For our selection criteria, we require candidates to have
$J_{125}-H_{160}$ colors redder than 1.2 AB mag, to be undetected
($<2\sigma$) in all imaging observations blueward of the $J_{125}$
band, and to be detected at $\gtrsim$5$\sigma$ detection in the
$H_{160}$-band.  This significance measurement in the $H_{160}$ band
is made in a small ($0.26''$-diameter) circular aperture to ensure the
result is more robust.  Significance measurements in scalable
apertures, with variable position angles and axis ratios, are more
susceptible to biases, since the apertures chosen to evaluate the
significance of sources are affected by noise within those sources
themselves.

In addition, our $J_{125}$-dropout candidates must not be detected at
$>$1.5$\sigma$ in more than one band blueward of the $J_{125}$ band.
Finally, to ensure that the optical imaging data are used to maximal effect
to reject low-redshift interlopers, we additionally require that candidates
have a $\chi^2$ value $<$2.5 in the
$B_{435}V_{606}i_{775}z_{850}Y_{105}$-band $\chi^2$ image.$^{39}$ The
$\chi^2$ image is equal to $\Sigma _{k} SGN(I_k) (I_k (x,y) / N_k)^2$.
$I_k (x,y)$ is the intensity of image $I_k$ at pixel $(x,y)$, $SGN(I_k)$ is
1 if $I_k>0$ and $-1$ if $I_k<0$, and $N_k$ is the RMS noise on that image.
In the definition of $\chi^2$, the index $k$ runs over all the relevant
images (i.e., those blueward of the dropout band).  

All of these requirements place very strong limits on the detected
optical light for our $z\sim10$ candidates and provide strong
discrimination against contamination by low-redshift sources.  While
we might occasionally reject some bona-fide $z\sim10$ candidates
utilizing this procedure, our primary objective with this selection is
to make it as clean as possible (given the inherent difficulty in
selecting $z\sim10$ galaxies at modest S/N levels).  We stress that
the volume densities we compute for $z\sim10$ candidates will not be
affected by this choice, since we also impose these same selection
criteria when computing the selection volumes.

We identified only one source which satisfied these stringent criteria
over the ultra-deep HUDF09 WFC3/IR data.  This source is listed in
Table~S\ref{tab:yjcandlist} and is shown in Figure 1 of the main text.
This candidate is detected at 5.4$\sigma$ in the $0.26''$-diameter
aperture used for selection and has an apparent magnitude of
28.9$\pm$0.2 mag.  The candidate has a significance of 5.8$\sigma$ in
a larger 0.35$''$-diameter aperture.\footnote{To be conservative we do
  not use the larger-aperture significance levels for our selection
  (since the larger aperture measurements can be more affected by
  uncertainties in the background level).} The candidate was not
present in two previous catalogs of $z\sim10$ candidates over the
WFC3/IR HUDF09 observations.  It falls just below 5$\sigma$ at
4.7$\sigma$ in these previous catalogs (which were based on only 50\%
of the two-year data set).

\begin{figure}
\epsscale{0.85}
\plotone{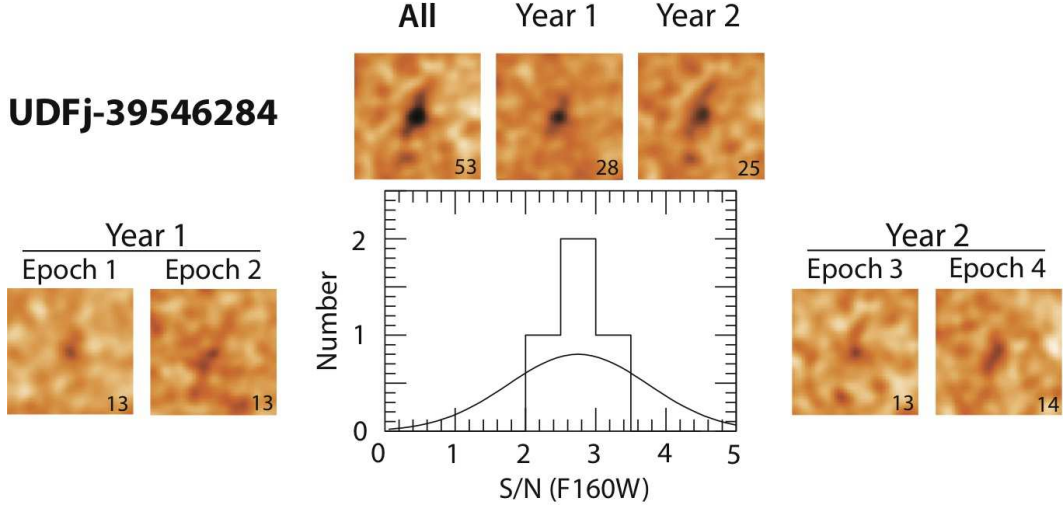}
\caption{Images of the single $z\sim10$ $J_{125}$-dropout candidate
  UDFj-39546284 from various subsets of the ultra-deep WFC3/IR
  $H_{160}$-band observations over the HUDF (53 orbits in total),
  including 13-orbit subsets of the ultra-deep in the lower panels.
  The distribution of the S/N levels for our $z\sim10$ candidate in
  the 13-orbit subsets of the $H_{160}$-band observations is also
  shown.  This S/N distribution for these 13-orbit subsets is expected
  to be approximately normal, with a mean of $5.4\sigma / 4^{0.5} \sim
  2.7\sigma$ and standard deviation of 1 (solid line).  The observed
  S/N distribution is in reasonable agreement with expectations (see
  also Figure~S\ref{fig:splits}).  The number of orbits of observations for
  each subset is given in the lower right corner.  Two orbits from the
  first-year observations are incorporated into the epoch 3
  observations to ensure each epoch had approximately the same
  13-orbit exposure time.  Note that our single $J_{125}$-dropout
  candidate is visible at $\geq2$$\sigma$ in all four 13-orbit subsets
  of the data and that the S/N of the object in each subset is
  consistent with the expected value of 2.7$\sigma$.
\label{fig:epoch}}
\end{figure}

Our second and third best candidates are $<$5$\sigma$ detections in
the $H_{160}$-band and hence do not make it into our election.  Our
second best candidate UDFj-38116243 (03:32:38.11,$-$27:46:24.3) is
detected at $4.9\sigma$ in the two-year observations, while our third
best candidate UDFj-39537198 (03:32:39.54,$-$27:47:19.9) is detected
at $4.0\sigma$.  It is interesting that our second best candidate
UDFj-38116243 in the full two-year observations was our most
significant candidate in the first-year observations. However it is
striking that its S/N is dramatically lower in the second year
observations (just 1.3$\sigma$) and so it is not a particularly
credible source.  It is discussed further in Appendix A.  It is also
noteworthy that our best candidate UDFj-39537198 (the only source in
our current $z\sim10$ sample) has both a substantially higher
significance level than the next most significant candidates, and a
much more uniform distribution of detection significances with time
(see Figure~S\ref{fig:epoch} and \S7).  As we will argue in \S7, this
strongly suggests that UDFj-39546284 corresponds to a real galaxy and
is not spurious.

To ascertain the nature of the new single $J_{125}$-dropout candidate,
we also examined the available IRAC $3.6\mu$ and $4.5\mu$ data. These
data reach to $\lesssim$27.7 AB mag and $\lesssim$27.1 mag,
respectively, at $1\sigma$.  After carefully modelling and subtracting
the IRAC flux from nearby neighbors,$^{40}$ no significant
($>$2$\sigma$ or $>$1$\sigma$) $3.6\mu$ or $4.5\mu$ detections are
found in the IRAC data (Figure 1 of the main paper).  This implies a
$H_{160}-3.6\mu$ color of $<$1.2 mag ($1\sigma$ limit) -- which is
more consistent with this source being a star-forming galaxy at
$z$$\sim$10 than an evolved or dusty galaxy at $z\sim2$.

\begin{deluxetable}{cccccccc}
\tabletypesize{\small}
\tablewidth{16.5cm} \tablecolumns{8} 
\tablecaption{Our single $z$$\sim$10 $J_{125}$-dropout candidate and
  three $z$$\sim$8.5-9 $Y_{105}$-dropout candidates identified over the
  ultra-deep HUDF09 WFC3/IR observations over the
  HUDF\label{tab:yjcandlist}}
\tablehead{ \colhead{Object ID} &
\colhead{R.A.} & \colhead{Dec} & \colhead{$H_{160}$\tablenotemark{a}} &
\colhead{$Y_{105}-J_{125}$\tablenotemark{b}} & \colhead{$J_{125}-H_{160}$} & \colhead{$r_{hl}$\tablenotemark{c}} & \colhead{$z_{est}$\tablenotemark{d}}}
\startdata
UDFj-39546284 & 03:32:39.54 & $-$27:46:28.4 & 28.92$\pm$0.18 & --- & $>$2.0 & 0.13$''$ & 10.3\\
\hline
UDFy-38135539 & 03:32:38.13 & $-$27:45:53.9 & 27.80$\pm$0.08 & 1.8$\pm$0.7 & 0.2$\pm$0.1 & 0.18$''$ & 8.7\\
UDFy-37796000 & 03:32:37.79 & $-$27:46:00.0 & 28.01$\pm$0.11 & $>$2.3 & -0.1$\pm$0.1 & 0.19$''$ & 8.5\\
UDFy-33436598 & 03:32:33.43 & $-$27:46:59.8 & 28.93$\pm$0.18 & $>$1.7 & 0.0$\pm$0.2 & 0.16$''$ & 8.6\\
\enddata 
\tablenotetext{a}{The magnitudes quoted here are based upon the light
  inside our large scalable apertures (and also include $\sim$0.2-0.3
  mag corrections for light on the wings of the PSF).  As such, they
  are significantly brighter than those quoted for our candidates in
  smaller apertures (e.g., in Figure S1).}
\tablenotetext{b}{ \textrm{L}ower limits on the measured colors are the  $1\sigma$ limits.}
\tablenotetext{c}{The quoted half-light radii are as observed and are
  not corrected for the PSF.  The half-light radius $r_{hl}$ for the
  PSF is 0.09$''$.}
\tablenotetext{d}{Estimated redshift. See Figure 2 of the main paper for
the redshift distributions}
\end{deluxetable}

\section{$Y_{105}$-dropout Selection}

\begin{figure}
\epsscale{0.95}
\plotone{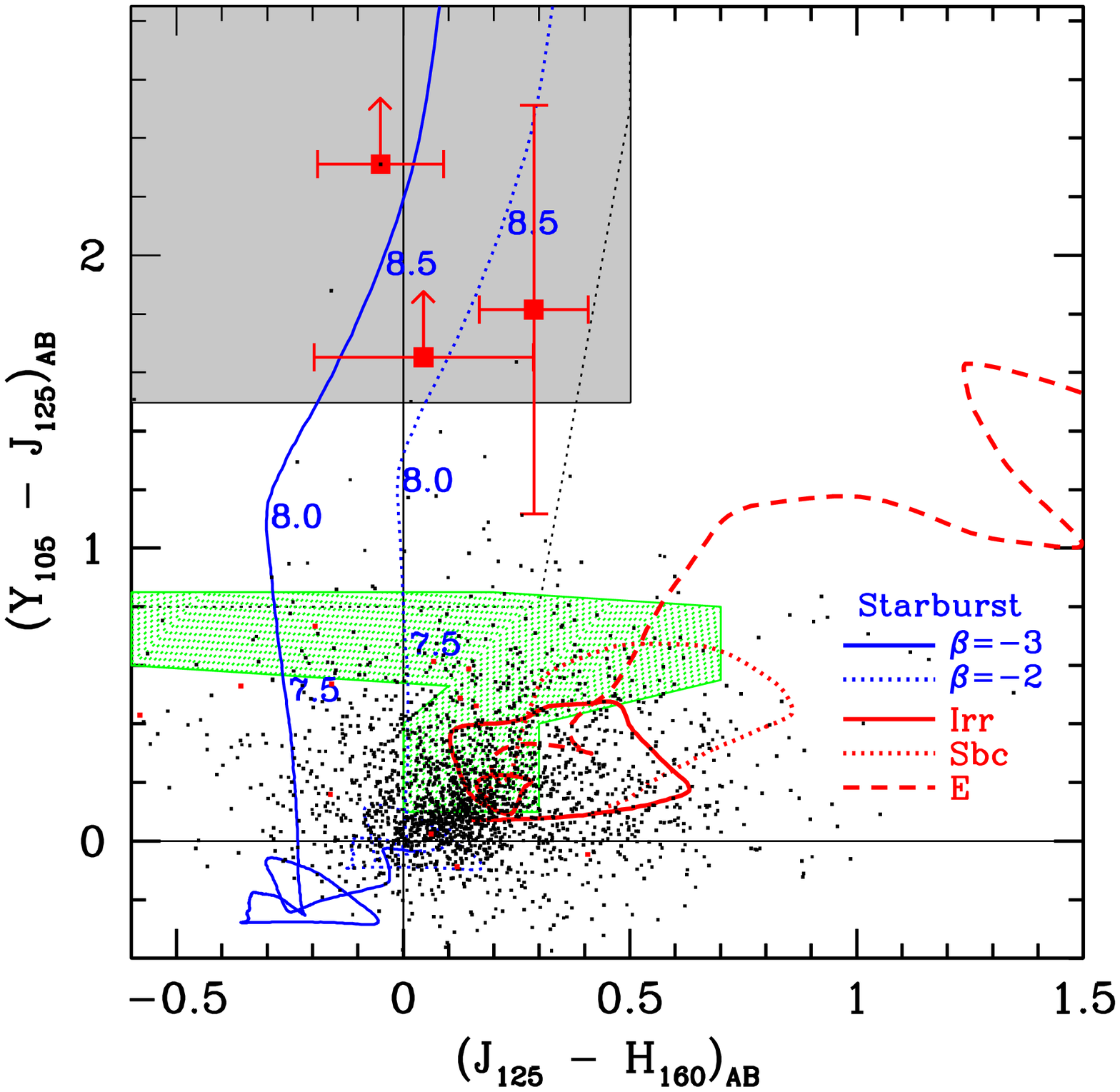}
\caption{$Y_{105}-J_{125}$ / $J_{125}-H_{160}$ two color selection
  used to identify $z\gtrsim8.2$ $Y_{105}$-dropout galaxies.  For
  context, the colors of star-forming galaxies with different
  $UV$-continuum slopes $\beta$ of $-3$, $-2$ versus redshift are also
  shown.  Also presented are the colors of low-redshift galaxy
  SEDs$^{41}$ over the redshift range $z\sim0$-2.  The
  $Y_{105}$-dropout selection window is presented in gray.  The
  objective of the current $Y_{105}$-dropout selection is to identify
  those $Y_{105}$-dropouts from the HUDF with the highest redshifts.
  The current sample of $Y_{105}$-dropouts is nonetheless a subset of
  the $Y_{105}$-dropouts identified in the Bouwens et al.$^3$ and
  Bouwens et al.$^9$ $Y_{105}$-dropout selections (the dotted black
  lines show the selection used by Bouwens et al.$^3$).  The hatched
  green region indicates the region in color space we would expect L,T
  dwarfs to lie.$^{42}$ \label{fig:sel}}
\end{figure}

We also carried out a search for galaxies at redshifts somewhat higher
than 8 (comparable to or higher than the redshift of the gamma-ray
burst source found$^{10,11}$ at $z\sim8.2$).  For this, we require
sources to have a very strong $Y_{105}-J_{125}>1.5$ breaks (see
Figure~S\ref{fig:sel}), to be undetected ($<2\sigma$) in all bands
$BViz$ blueward of the break, and to be detected at
$\gtrsim$5.5$\sigma$ in the $J_{125}$-band to ensure they correspond
to real sources.  This $Y_{105}$-dropout selection criterion is more
stringent than what we applied in our earlier searches over the
ultra-deep HUDF WFC3/IR data$^{3,9}$ and identifies galaxies in the
range $z$$\sim$8.2-8.8, with a mean redshift of $z\sim8.5$ (see Figure
2 of the main text).

\begin{figure}
\epsscale{0.88}
\plotone{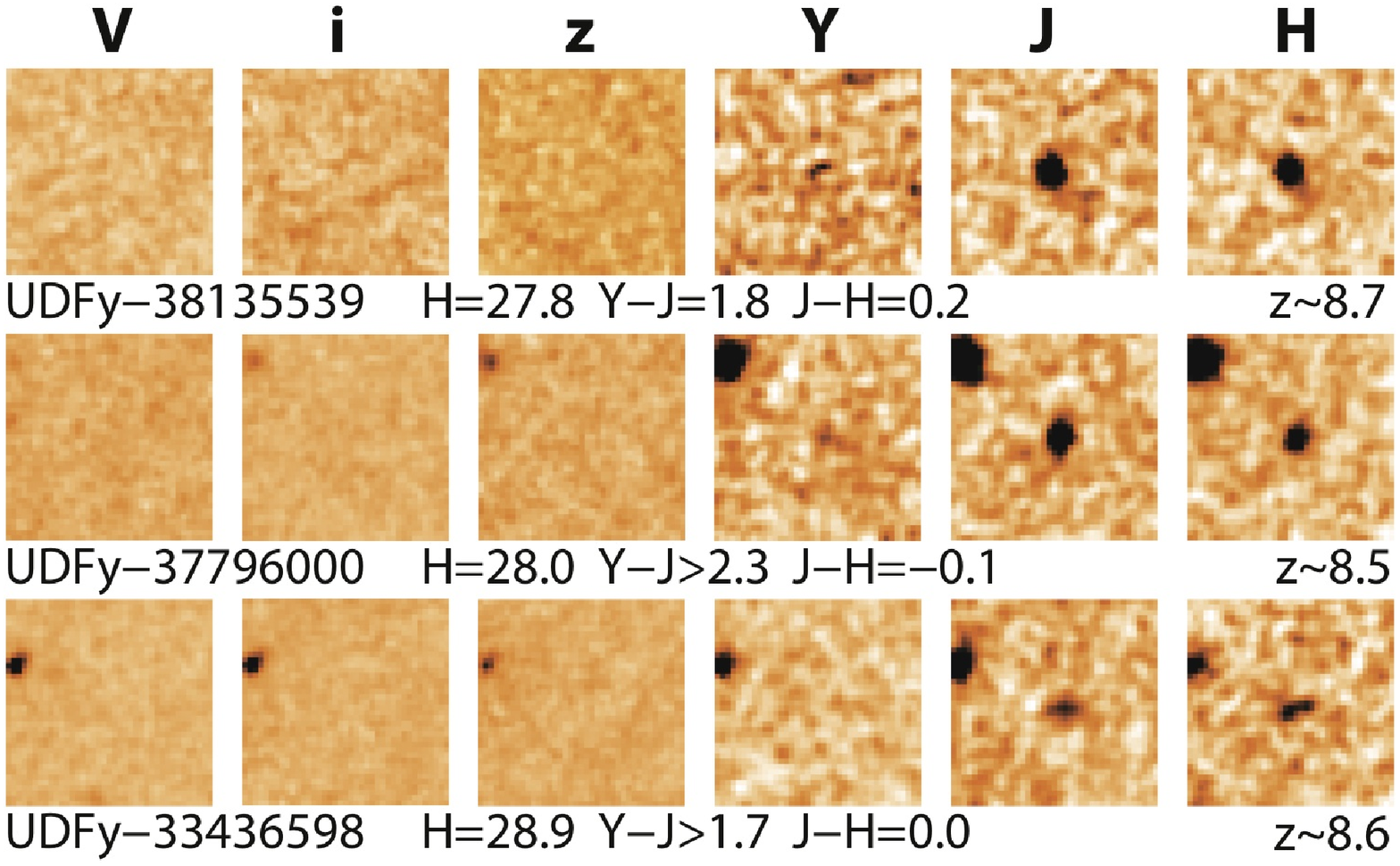}
\caption{$V_{606}i_{775}z_{850}Y_{105}J_{125}H_{160}$ image cutouts of
  three $z$$\sim$8.5 $Y_{105}$-dropout galaxy candidates we identified
  in the ultra-deep HUDF WFC3/IR observations (\S5).  None of the
  candidates are detected in the deep ACS $BViz$ observations. The
  candidate UDFy-38135539 has a tentative spectroscopic confirmation
  at $z\sim8.6$.$^{14}$ Other properties of the candidates are given
  in Table~S\ref{tab:yjcandlist}.  Each cutout is $2.4''\times2.4''$
  on a side and with North up.\label{fig:cutouts}}
\end{figure}

We identified three sources which satisfied these selection criteria.
These sources are listed in Table~S\ref{tab:yjcandlist} and shown in
Figure~S\ref{fig:cutouts}.  Two of our three candidates (the brightest
ones) were already presented in the Bouwens et al.$^3$
$Y_{105}$-dropout selection, the McLure et al.$^4$ $z$$\sim$6-9
search, the Bunker et al.$^5$ $Y_{105}$-dropout selection, the Yan et
al.$^6$ $Y_{105}$-dropout sample, and the Finkelstein et al.$^7$
$z$$\sim$6.3-8.6 sample.

\clearpage

\section{$J_{125}$-dropout selection over the Early Release Science 
observations}

To constrain the volume density of sources at the bright end of
$z\sim10$ LF, we also conducted a search for $z$$\sim$10
$J_{125}$-dropout candidates over the wide-area ERS observations.  The
ERS observations cover $\sim$10$\times$ as much area as probed by our
ultra-deep WFC3/IR observations over the HUDF.  This large area is
very valuable for constraining the volume density of the likely
intrinsically very rare sources at the bright end of the $z\sim10$ LF.

Both our catalog construction procedure and selection criteria remain
the same as what we used for the ultra-deep HUDF09 field.  We find no
candidates with $J_{125}-H_{160}$ colors $>$1.2 and optical
non-detections over the ERS observations.  The only sources that
satisfy those criteria were also quite bright at $5.8\mu$ and
$8.0\mu$, and hence were likely $z$$\sim$2-3 dusty galaxies (having
$H_{160}-5.8\mu$ colors redder than $\sim$2).

\begin{figure}
\epsscale{0.85}
\plotone{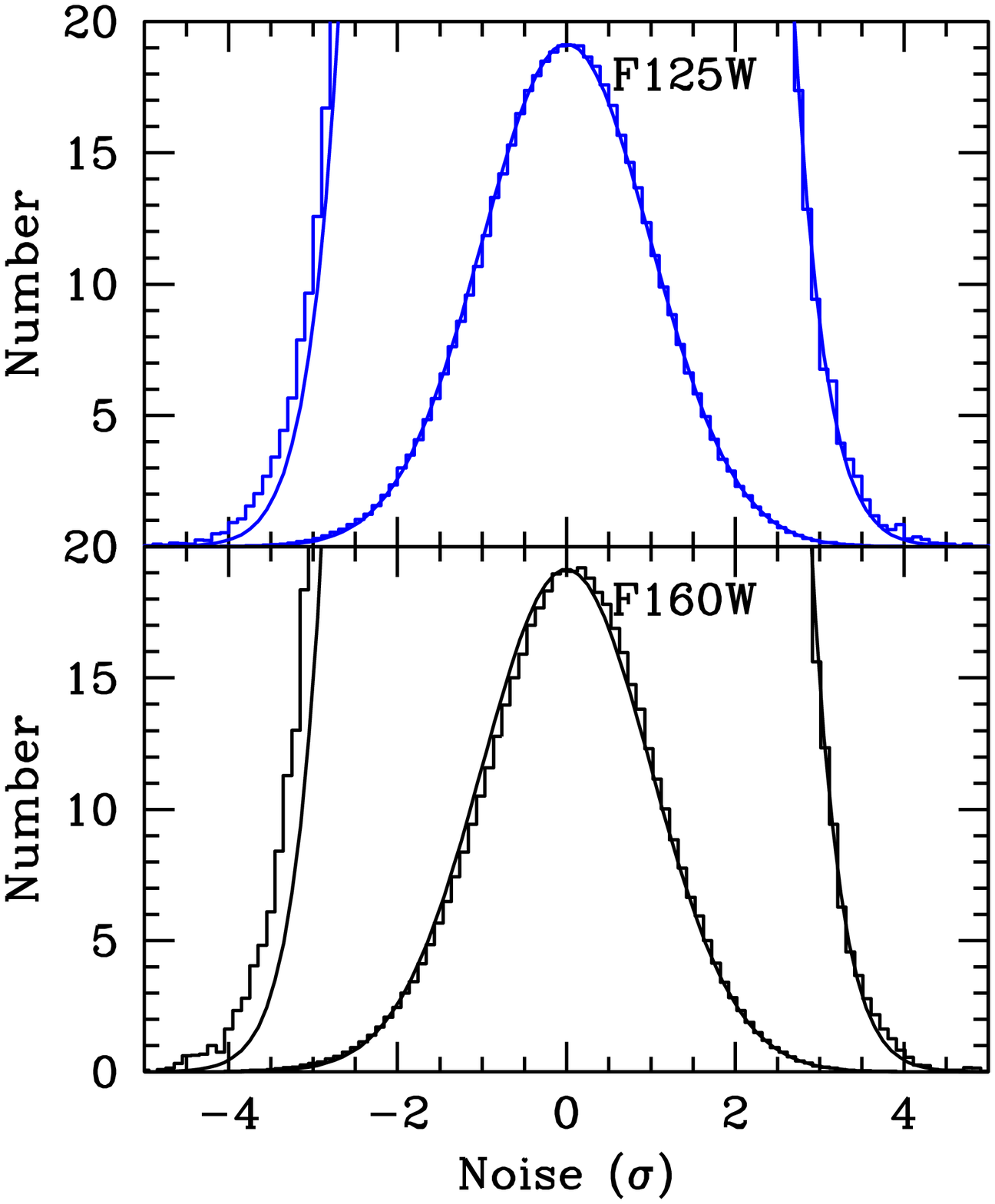}
\caption{Observed distribution of noise fluctuations in 3 pixel by 3
  pixel regions in the F125W and F160W band observations (histograms:
  upper and lower panels, respectively).  The solid line shows the
  model noise distribution, for perfectly Gaussian noise.  While the
  observed distribution of fluctuations does show slight skewness,
  this distribution is almost perfectly Gaussian out to significance
  levels of $\sim$3.5$\sigma$.  The noise distribution for positive
  fluctuations appears to only exceed that expected from normal
  statistics, by less than a factor of 2.\label{fig:noise}}
\end{figure}

\begin{figure}
\epsscale{0.85}
\plotone{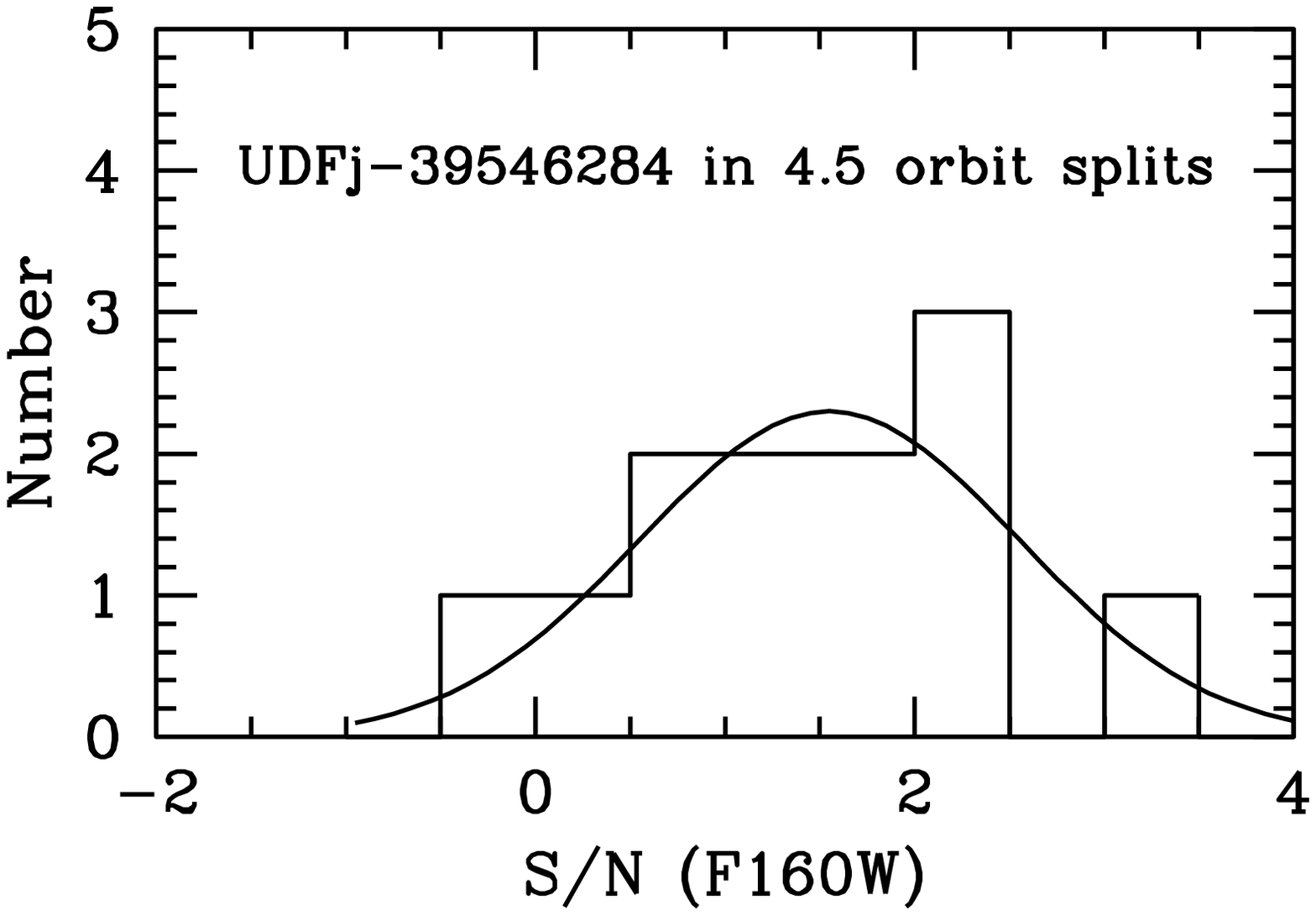}
\caption{The distribution of S/N levels for our $z\sim10$ candidate in
  12 independent 4.5-orbit splits of the $H_{160}$-band observations
  (53 orbits in total).  Because 53 orbits cannot be split evenly into
  12 4.5-orbit subsets, one split only included 3.5 orbits.  This
  distribution appears to be a good fit to a normal distribution with
  a mean of 1.6$\sigma$ and standard deviation of 1$\sigma$.  A mean
  of 1.6$\sigma$ is exactly what one would predict based upon a
  scaling of the S/N with the square root of the number of exposures,
  i.e., $5.4\sigma / 12^{1/2}\sim1.6\sigma$.  The general agreement of
  the S/N distribution for our $J_{125}$-dropout candidate with that
  expected for Gaussian noise suggests that this candidate is not
  spurious.\label{fig:splits}}
\end{figure}

\section{Possible Contamination} 

\textit{Contamination from Spurious Sources:} Perhaps the most
important issue to examine in assessing the present $J_{125}$-dropout
selection is determining whether our candidate is real or whether it
could be due to random fluctuations in the noise (i.e., spurious).
Since our candidate is only detected in a single passband at
$\sim$5.4$\sigma$, one must consider the possibility that it could be
spurious.

\textit{Contamination from Spurious Sources: Noise Distribution:} To
determine the probability that our candidate could be spurious, we
first examine the noise distribution in our ultra-deep WFC3/IR images.
This is important since it will allow us to quantify the extent to
which noise in WFC3/IR images has non-Gaussian characteristics and
therefore might allow for $\gtrsim$5$\sigma$ noise fluctuations (i.e.,
outliers) simply by chance.

We start by examining the noise distribution in the $J_{125}$-band
image, since we would not expect many faint sources with emission in
the $J_{125}$-band alone and not in the other frames.  The
distribution of flux values in 3 pixel-by-3 pixel boxcar smoothed
regions is shown in the upper panel of Figure~S\ref{fig:noise} after
masking out all those regions which show $\geq1.5\sigma$ detections in
a coaddition of the optical, $Y_{105}$, and $H_{160}$ images.  A
similar noise distribution is shown in the lower panel for
fluctuations in the $H_{160}$-band, after masking out those regions
detected in the other bands.  Overall, the noise distributions appear
to be approximately Gaussian to the extent that it can be tested with
our data and only show minor deviations.  This suggests that apparent
$5\sigma$ events are indeed significant at approximately $5\sigma$,
and therefore the $5.4\sigma$ $J_{125}$-dropout in our sample is not
likely a chance noise fluctuation.

\textit{Contamination from Spurious Sources: Splitting the Data:}
Another procedure for testing the reality of our $J_{125}$-dropout
candidate is to split the $H_{160}$-band data into four independent
subsets of 13 orbits (representing $\sim$25\% of the total data set)
and evaluate the significance of this candidate in each subset.  The
four $\sim$13-orbit splits are shown in Figure~S\ref{fig:epoch}, and
it is evident that our candidate is significant in each split.  In a
$0.26''$-diameter aperture, it is significant at 3.4$\sigma$,
2.7$\sigma$, 2.3$\sigma$, and 2.7$\sigma$, with a mean of 2.8$\sigma$.
The significance is thus $\geq$2.3$\sigma$ in all four splits (see
Figure~S\ref{fig:splits}). Since 2.3$\sigma$ events occur with
probability $\lesssim$1\% and there are only
$\sim(133''/0.2'')^2\sim700^2$ independent resolution elements over
the WFC3/IR HUDF in which such sources could be found, there is
$\lesssim$1\% chance that this event would be found at random.  A
modification of this test to consider even smaller 4.5-orbit subsets
of the data provides a more rigorous assessment of the robustness of
our candidate against non-Gaussian fluctuations.  The distribution of
significance levels for our candidate in 4.5-orbit splits is
consistent with a normal distribution with a mean of 1.6$\sigma$ and
scatter of 1$\sigma$ (Figure~S\ref{fig:splits}).

A slight variation of the above tests would be to see if the candidate
were amongst the most significant candidates in the first and second
year observations.  We would not expect there to be a highly
significant candidate at approximately the same position in both the
first and second year observations if it were spurious.  Constructing
$J_{125}$-dropout catalogs from the first and second year observations
and extending our selection down to $3\sigma$, we found a
$\geq3\sigma$ $J_{125}$-dropout candidate in the second-year
observations within $0.06''$ of a similar $\geq3\sigma$
$J_{125}$-dropout candidate in the first-year observations.  This
candidate was the 2nd most significant $J_{125}$-dropout candidate in
the first year observations and the 10th most significant in the
second.  The probability that one of the ten most significant
candidates in both the first and second year observations would occur
within $0.06''$ of each other is
$\sim10^2(\pi(0.06'')^2)/(133'')^2\sim0.01$\%.

\textit{Contamination from Spurious Sources: Negative Image Test:} Two
other procedures for assessing the probability that our candidate
might be spurious are the ``negative image'' test$^{35}$ and a test
where we look for similar $5\sigma$ detections (single or multiple
band) in bluer filters.  The rationale for using the negative images
is that the noise properties are fairly symmetric in the positive and
negative directions (see Figure~S\ref{fig:noise}), and therefore a
search for $J_{125}$-dropout candidates in negative images allows us
to assess the relevance of spurious sources to our selection.  A
search for $J_{125}$-dropout candidates on the negative $H_{160}$-band
images yielded no candidates.  Nonetheless, we do note the presence of
a source just below our selection threshold (at 4.8$\sigma$: but no
others above 4.4$\sigma$).  Though it is difficult to calculate
precisely how much more significant our 5.4$\sigma$ $J_{125}$-dropout
candidate is than this 4.8$\sigma$ spurious source, for Gaussian
statistics it is approximately $\sim$20$\times$ more likely.  This
again suggests our selection is probably free of contamination from
spurious sources.

\textit{Contamination from Spurious Sources:
  $Y_{105}+J_{125}$/single-band $J_{125}$ test:} A search for sources
which were $5\sigma$ detections in a 53-orbit $Y_{105}+J_{125}$ image
(the two-bands are combined to construct an image with an equivalent
number of exposures to our two-year $H_{160}$-band stack and hence
similar noise statistics) revealed no $5\sigma$ sources with
$H_{160}-Y_{105}/J_{125}$ colors $>$1.2 not seen in the
$H_{160}$-band.  The most significant source in the $Y_{105}+J_{125}$
image was 4.5$\sigma$.  Again this suggests that our sole $z\sim10$
candidate is not spurious.  However, we do note that we do find a
$5.1\sigma$ single-band detection if we examine only the 34-orbit
$J_{125}$-band (and not include the $Y_{105}$-band in the stack).
While the existence of such a source in the $J_{125}$-band data could
indicate that non-Gaussianity is still important for $\sim$30-orbit
exposures with WFC3/IR,\footnote{It is possible that non-Gaussianity
  may explain the existence of UDFj-38116243, one of our
  $J_{125}$-dropout candidates and apparent $5\sigma$ detection in the
  first-year data (see Appendix).} it seems clear that non-Gaussianity
will be less important for our 53-orbit $H_{160}$-band stack.

Finally, we remark that we identified our sole $J_{125}$-dropout
candidate as the most promising candidate shortly after the
second-year WFC3/IR $H_{160}$-band observations over the HUDF began
and when there were 15 orbits of $H_{160}$-band observations
remaining.  The subsequent observations were therefore a test of the
reality of the candidate.  As is evident in Figure~S\ref{fig:epoch},
the source is significant at $\geq$2.5$\sigma$ in the observations
taken after this point -- which would occur at only 1\% probability if
the source were spurious.

Most of the above tests suggest contamination from spurious sources at
$\sim$1\% probability -- though the results from the negative image
test and $Y_{105}+J_{125}$/$J_{125}$-band tests suggest a
contamination rate as high as 10\%.  To be conservative, we will
assume that the contamination rate from spurious sources in our
selection is 10\%, or $\sim$0.1.  However, we stress that such formal
analyses should be treated with caution as unknown effects could be
present in the data.

\textit{Contamination from Red Low-Redshift Galaxies:} An equally
important concern for the present $J_{125}$-dropout selection is
contamination from compact, low-surface brightness, low-redshift, very
red sources (typically because of age or dust).  It is plausible that
such sources could satisfy our $J_{125}$-dropout selection criteria,
with $J_{125}-H_{160}$ colors redder than 1.2 and showing very little
flux in the ultra-deep optical data.  Obviously, we might expect such
sources to be somewhat brighter at redder wavelengths, and so the IRAC
data provide us with a way of constraining this possibility.  As we
noted in the previous section, each of our candidates is undetected in
the $3.6\mu$ band, allowing us to set a $H_{160}-3.6\mu<1.2$ limit on
its color blueward of the $J_{125}-H_{160}$ break -- providing some
evidence that $J_{125}$-dropouts in our selection are not just red
galaxies at $z\sim2$.

An independent way of examining whether lower-redshift red galaxies
are a concern for our $J_{125}$-dropout selection is performing an
analogous selection for $z\sim8$ $Y_{105}$-dropouts -- assuming that
no $H_{160}$-band observations are available.  We therefore select
$Y_{105}$-dropouts using two criteria: being undetected in the optical
data and satisfying a single color $Y_{105}-J_{125}>1.2$ cut (and no
$J-H$ criterion).  In performing this experiment, we only remove
sources that are bright enough in the optical that they would be
detected if the source had the same $H_{160}$-band magnitudes as our
$J_{125}$-dropout candidates (which are typically $\sim$0.5 mag
fainter).  We identify exactly the same sources in this
$Y_{105}$-dropout selection as we would select also including a
$J_{125}-H_{160}<0.5$ selection on the $J-H$ color (8 sources in total
from the HUDF09 and HUDF09-2$^9$ fields).  This suggests that our lack
of ultra-deep imaging data redward of the $H_{160}$-band does not
substantially add to the contamination.

This test demonstrates how powerful an optical non-detection criterion
combined with a strong color break in broadband SED can be in
identifying $z\geq7$ galaxies, given the colors of faint sources in
deep field data.  It also suggests that other techniques that do not
take advantage of information about the actual prevalence of sources
with specific colors at faint magnitudes may actually overestimate the
contamination rate for high redshift selections (e.g., ``photometric
redshift'' techniques that compute $P(z)$ based upon $\chi^2$ fits to
various SED templates versus redshift).

This is a very powerful independent test that suggests our single-band
approach is much more robust than it might first appear.  The reason
that such a selection is so useful is that the HUDF has such a variety
of deep data in so many filters over a wide wavelength range to check
for contamination.

\textit{Contamination from Photometric Scatter:} Another possibly
significant source of contamination for the current $J_{125}$-dropout
selection are sources that satisfy our $J_{125}$-dropout criterion due
to noise in the photometry.  To assess this possibility, we considered
each faint source in our $H_{160}$-band catalog in turn, randomly
selected a source in $H_{160}$-band magnitude range 26.5 to 27.5 AB
mag (taken to be representative of the color distribution at fainter
magnitudes), rescaled their fluxes to match the fainter source in our
$H_{160}$-band catalog, added noise, and finally reapplied our
$J_{125}$-dropout selection criteria.  Repeating this simulation many
times, we found that $\sim$0.1 contaminants per field satisfy our
$J_{125}$-dropout criterion, simply as a result of noise.  A detailed
description of this procedure (but for $z$ or $Y$ dropouts) is
discussed in Bouwens et al.$^9$

\textit{Other Sources of Contamination:} Other possible sources of
contamination for our $J_{125}$-dropout includes transient sources
like SNe or stars.  Contamination by SNe seems extremely unlikely
since the $J_{125}$ and $H_{160}$ observations were taken within 5
days of each other.  Contamination by stars also seem rather
implausible, since the source appears to be slightly resolved, and the
only stars which are red enough to satisfy the $J_{125}-H_{160}>1.2$
criterion are extreme carbon stars or Mira variables$^{44}$ -- which
would need to be located well outside our galaxy$^{45}$ given their
luminosity.

Lastly, all of our candidates are sufficiently separated from extended
foreground sources that they are unlikely subcomponents of these
sources.  A somewhat parallel concern in doing candidate selection
near extended sources is that the detection significance of possible
sources is enhanced by light from the extended wings.  It seems quite
probable that an independent selection of $z$$\sim$8-10 candidates$^6$
may have been subject to significant contamination as a result of this
issue (see arXiv:0912.4263 for a discussion of some problems with that
selection).

\textit{Summary:} The above simulations suggest we would find
$\sim$0.2 contaminants in the $z$$\sim$10 $J_{125}$-dropout selection
we perform over the HUDF09 observations: $\sim$0.1 from spurious
sources and $\sim$0.1 from photometric scatter.  This implies that 0.8
$J_{125}$-dropout candidates from our selection correspond to
$z\sim10$ galaxies.  Of course, it is important to remember that such
simulations can often have significant uncertainties and so we will
also consider the case that we have not detected any $z\sim10$
sources.

In the sections which follow, we will explore the implications (1) if
0.8 $J_{125}$-dropout candidates here are at $z$$\sim$10 and (2) if
none are.

\textit{Contamination in $Y_{105}$-dropout selections:} The only
important source of contamination for $Y_{105}$-dropout selections
seems to be sources that enter the selection as a result of
photometric scatter (i.e. noise),$^3$ and even for this source of
contamination, we derive a contamination rate of $<4\%$ from
Monte-Carlo simulations.

\section{Contrasting the current $z\sim 10$ selection with previous 
$z\sim 10$ selections}

The second year of WFC3/IR observations do not provide significant
support for the reality of the three candidates we initially
identified in the first-year of observations, as we describe in
Appendix A.  Should we have more confidence in the sole $z\sim10$
$J_{125}$-dropout candidate found in the full two-year observations
than we had in our $z\sim10$ candidates from the first-year
observations?  There are at least four reasons why we would argue the
answer is \textit{yes}.

First, we now have approximately twice the total exposure time to
evaluate the reality of our $z\sim10$ candidates, as we had
previously.  This additional exposure time is valuable since it
provides for a very rigorous test of the reality of any possible
candidates in our selection.  In the present case, our sole $z\sim10$
candidate shows up at $\geq2.3\sigma$, with a mean of 2.8$\sigma$, in
four $\sim$13-orbit splits (Figure~S\ref{fig:epoch}) and with a mean
of 1.6$\sigma$ in 12 $\sim$4.5-orbit splits
(Figure~S\ref{fig:splits}).  The increased number of exposures also
results in the candidate showing a higher significance level than any
of the previous candidates.  Lastly, the larger number of exposures
results in the noise being more Gaussian in nature (from the
well-known central limit theorem in statistics), making apparent high
$\sigma$ events genuinely more significant.

Second, we now use much more conservative selection criteria than we
used in our first-year selection.  The fixed apertures we utilize to
evaluate the significance of sources are much less susceptible to
biases than the scalable Kron-like apertures we had previously used,
and only one of our previous three candidates would have been selected
with such a procedure.

Third, our current $z\sim10$ candidate is much more unique in terms of
its overall significance levels than the candidates in the first-year
selection.  While there were a substantial number of sources between
$4\sigma$ and $5\sigma$ in our first-year catalog, this contrasts with
only two sources in that significance range here.  This argues that
the candidate is much more unique overall and it is not produced by
sources scattering into our selection through noise.

Fourth, our current $z\sim10$ candidate is found at magnitude levels
much more consistent with it corresponding to bona-fide high-redshift
galaxies than was the case for the first-year candidates.  The reason
for this is simple.  Due to rapid changes in completeness, one expects
candidates (in general) only to begin appearing a few tenths of the
magnitude brightward of the selection limit since it is only at such
magnitudes that the completeness will be high enough to observe high
redshift galaxies in significant numbers.  Below the selection limit,
the number of bona-fide sources becomes very small, first because the
search is becoming seriously incomplete, and second because the
contamination rate can increase rapidly.  This is illustrated in the
specific case of our $z\sim10$ searches in Figure~S\ref{fig:counts}
and in Figure~S\ref{fig:numc}. The current $z\sim10$ candidate is
several tenths of a magnitude brighter relative to the selection limit
($\sim$29.8 AB mag in Figure 1) than were the three first-year
candidates.  This contrasts to the first year candidates in the same
apertures which were actually $<$5$\sigma$, and in two cases, much
less (see Appendix A).  The likelihood of contamination for the
present $z\sim10$ candidate (i.e., that the source is spurious) is
therefore quite a bit smaller.

Together these improved tests and cross-checks, the higher
significance and more robust S/N, all suggest that the new source is
more likely to be a high-redshift galaxy -- and, combined with the
photometric data, a plausible $z\sim10$ candidate.  It is also
reassuring that the surface density implied by this single candidate
is close to what we would expect extrapolating from lower redshift
(see \S9 and Figure~S\ref{fig:numc}).

\section{Expected Numbers} 

To provide a baseline for interpreting the number of high-redshift
dropout candidates we have found over our search fields, we compare
the present findings with what we might expect extrapolating
lower-redshift results to $z\sim10$ assuming no evolution. 

We use a similar procedure as described in previous work$^{3,21}$ to
account for the many different selection effects (and incompleteness).
We start by generating catalogues based upon the model LFs, creating
realistic pixel-by-pixel simulations of the model sources, adding
these simulated images to the actual ACS+WFC3/IR data, and then
processing the images (and selecting the sources) in the same way as
on the real data.  By adding the model sources to the actual imaging
data, we naturally account for the effect that foreground sources have
on both the completeness of our dropout selections and the accuracy of
our photometry.$^1$ 

The spatial profiles of the sources in the simulations are modelled
using the pixel-by-pixel morphologies of similar-luminosity $z\sim4$
$B$-dropout galaxies from the HUDF$^1$ scaled in size using the
$(1+z)^{-1}$ size-redshift relation.$^{46-48}$ The sources are also
taken to have a $UV$-continuum slope distribution with a mean $\beta$
equal to $-3$ and a scatter of 0.4 to match the observed trends with
redshift.$^{49}$

\begin{figure}
\epsscale{0.6}
\plotone{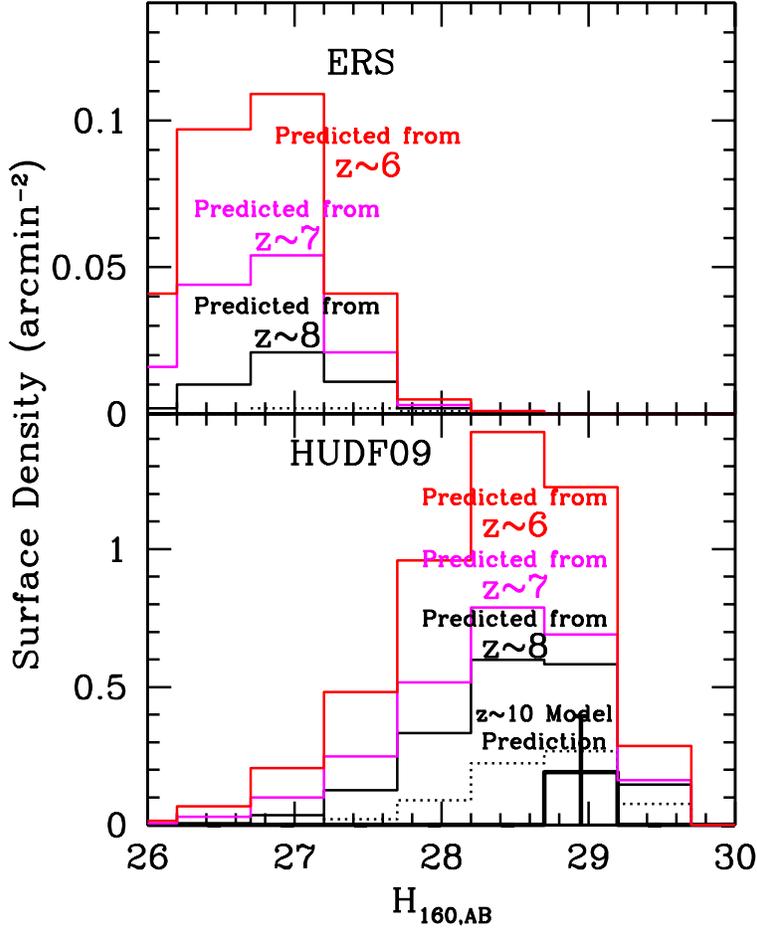}
\caption{Observed surface densities of $z$$\sim$10 $J_{125}$-dropout
  candidates (\textit{histograms shown with thick black lines}) in the
  wide-area ERS observations (\textit{top panel}) and ultra-deep
  HUDF09 observations (\textit{bottom panel}).  The observed surface
  densities have been corrected assuming $\sim$20\% contamination in
  the HUDF09 observations (see \S7).  For comparison, we have also
  plotted the expected surface densities of $z\sim10$
  $J_{125}$-dropout galaxies assuming no evolution in the UV LF$^{1}$
  from $z\sim6$ (\textit{red histogram}), $z\sim7$ (\textit{magenta
    histogram})$^{9}$, and $z\sim8$ (\textit{black histogram}).  The
  $z\sim8$ LF we use for this no-evolution calculation adopts an
  $M^*\sim-19.5$, $\phi^*\sim0.0014$ Mpc$^{-3}$, and $\alpha=-1.74$
  (obtained using the Bouwens et al.$^{22}$ LF-fitting formula while
  keeping $\alpha$ fixed, but we note that it is consistent with
  observations$^9$).  We also show the predicted $J_{125}$-dropout
  surface densities adopting a $z\sim10$ model LF (dotted histogram),
  with $M^*\sim-18.8$, $\phi^*\sim0.0016$ Mpc$^{-3}$, and
  $\alpha=-1.74$ (these parameters also come from the Bouwens et
  al.$^{22}$ fitting formula but again keeping $\alpha$ fixed).  The
  surface densities of $z\sim10$ galaxies expected at $\sim$29 AB mag
  are in good agreement with what we would infer if our single
  $J_{125}$-dropout candidate over the HUDF actually corresponds to a
  $z\sim10$ galaxy.  We refer the reader to e.g. Bouwens et
  al.$^{3,9}$ for a discussion of the surface densities for $z\sim8$
  $Y_{105}$-dropout selections from the HUDF09
  observations.\label{fig:numc}}
\end{figure}

For these no-evolution cases, we predict 23 and 12 $J_{125}$-dropouts
over the HUDF and ERS fields using the $z\sim6$ LF$^1$, 12 and 5
dropouts over the HUDF and ERS fields using the $z\sim7$ LF$^{9}$, and
9 and 2 dropouts over the HUDF and ERS fields using a $z\sim8$
LF$^{22}$ (consistent with the LF from Bouwens et al.$^{9}$).

Since the LF does evolve quite markedly with redshift,$^{22}$ we can
make a much more realistic prediction for our $z\sim10$ search
results.  We provide this prediction based upon the Bouwens et
al.$^{22}$ LF-fitting formula where the Schechter parameters are
$M_{UV,AB}=-18.8$ and $\phi^*=0.0016$ Mpc$^{-3}$.  We fixed $\alpha$
to the same value $-1.74$ found at $z\sim6$.$^1$ Based upon this LF,
we predict 3 $J_{125}$-dropouts over the HUDF at $z\sim10$ and 0.2
$J_{125}$-dropouts at $z\sim10$ over the ERS observations.

The predicted $J_{125}$-dropout surface densities are presented in
Figure~S\ref{fig:numc} for the various $UV$ LFs.  Not surprisingly,
significant numbers would be expected at brighter magnitudes based
upon the $UV$ LF at $z\sim6$ and $z\sim7$ if there is no evolution.
However, we only detect a candidate $z\sim10$ galaxy at the faint end
of our deep HUDF09 search (i.e., $H_{160,AB}\sim29$ for the HUDF09
observations).  This again points towards significant evolution in the
UV LF from $z\sim10$ to $z\sim7$.  In the HUDF, the surface density of
$J_{125}$-dropout candidates observed is comparable to that expected
from an extrapolation of the $UV$ LF to $z\sim10$.  It is remarkable
that the Bouwens et al.$^{22}$ LF fitting formula is plausibly
consistent with the observations all the way out to $z\sim10$.

We expect the field-to-field variance on the present $J_{125}$-dropout
searches to be 39\% and 40\% in our HUDF09 and ERS searches,
respectively -- assuming a redshift selection window with width
$\Delta z \sim 1.3$; a search area of 4.7 arcmin$^2$ and 39.2
arcmin$^2$, respectively; and bias factors of $\sim$8 and $\sim$12,
respectively.  These bias factors correspond to source volume
densities of $\sim4$$\times$10$^{-4}$ Mpc$^{-3}$ and
$\sim$3$\times$10$^{-5}$ Mpc$^{-3}$, respectively, expected to be
probed in our HUDF09 and ERS searches and were computed using the
Trenti \& Stiavelli$^{20}$ cosmic variance calculator.  While large,
these uncertainties are smaller than those expected, given the very
small numbers (and thus large Poissonian errors).

The predicted redshift distributions for our $J_{125}$-dropout selection is
given in Figure 2 of the main text.  For this calculation, we assume that
the $UV$ LF at $z\sim10$ has a $M_{UV} ^*=-18.8$, $\alpha=-1.74$, and
$\phi^*=0.0016$ Mpc$^{-3}$, i.e., using the Bouwens et al.$^{22}$ LF
fitting formula to extrapolate the $z\sim4-7$ LF results to $z\sim10$.  The
mean redshift for our $z\sim10$ $J_{125}$-dropout selection is 10.3.  Also
presented in Figure 2 of the main text is the predicted redshift
distribution of the present $Y_{105}$-dropout selection (assuming $M_{UV}
^*=-19.5$, $\alpha=-1.74$, and $\phi^*=0.0014$ Mpc$^{-3}$) for each
$z\sim8.5$ candidate.  The $J-H$ color for UDFy-38135539 is 0.4 mag 
redder than UDFy-37796000 and 0.3 mag redder than UDFy-33436598 --
suggesting probable redshifts of 8.7, 8.5, and 8.6, respectively for the
sources.  While the latter redshift estimate is somewhat higher ($\Delta
z$$\sim$0.2-0.3) than the McLure et al.$^4$ estimate for these two
$Y_{105}$-dropouts, the present estimate adopts a somewhat bluer
$UV$-continuum slope $\beta$ distribution (as suggested by ref 49) than the
likely SED templates used by McLure et al.$^4$ for their redshift
estimates.

\begin{figure}
\epsscale{0.85}
\includegraphics[angle=270,scale=0.7]{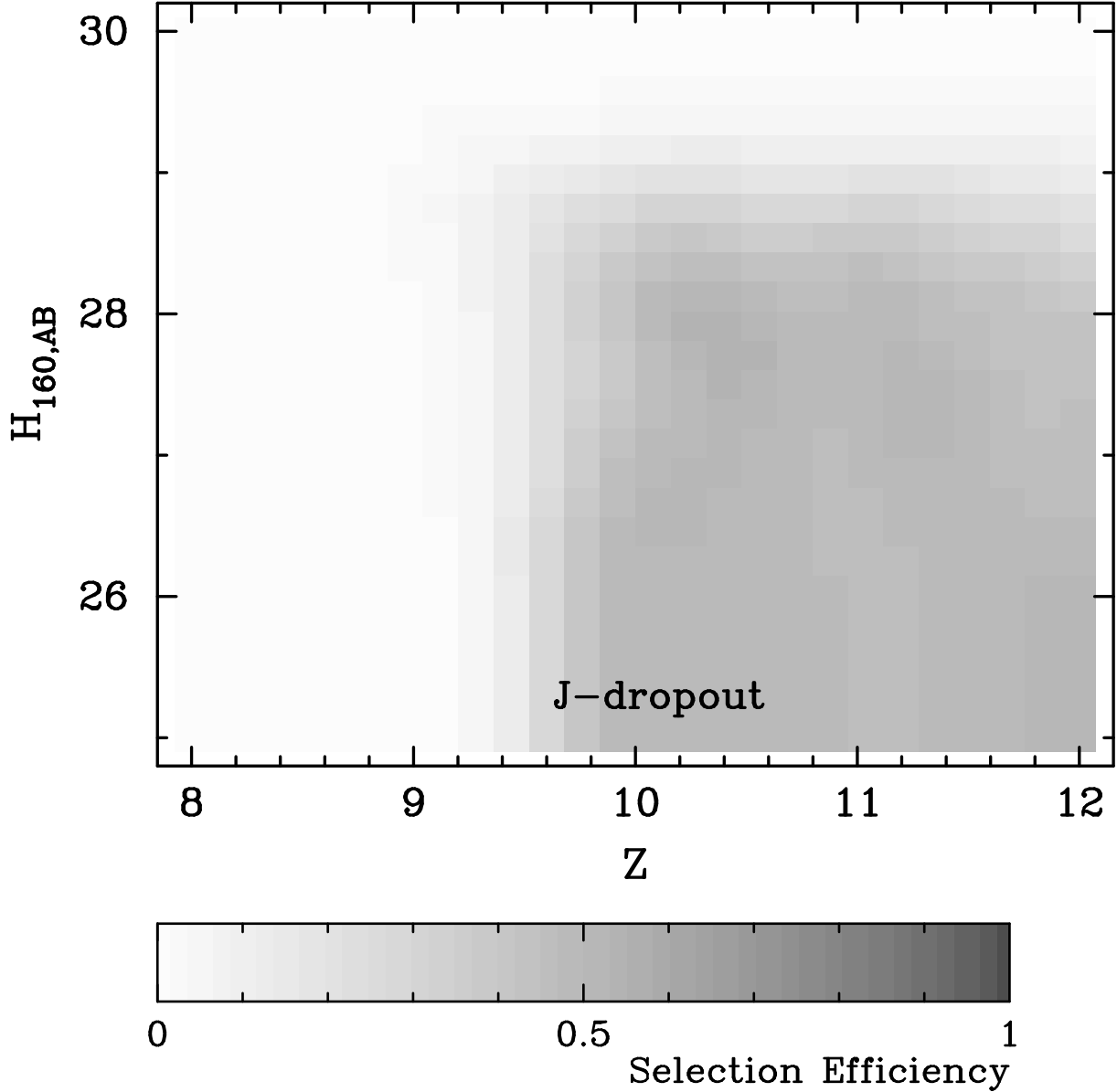}
\caption{The efficiency with which we select star-forming galaxies
  versus redshift at specific apparent $H_{160}$-band magnitudes in
  the HUDF09 observations.  These efficiencies are important for our
  LF constraints (\S10) and were estimated by adding model galaxies
  with realistic colors and sizes to the actual observations and
  attempting to reselect them using the same procedure as we use on
  the observations (see \S9).  The reason the selection efficiency is
  lower than 100\% is due to the area covered by both bright and faint
  foreground galaxies.  The faint foreground galaxies make it
  difficult to register a clear ``null'' detection in the optical.  We
  even lose $\sim$10\% of the area (even in empty regions) due to our
  requirement that $\chi^2$ (from the BVizY data) $<$2.5 and the fact
  that such a low $\chi^2$ can be exceeded simply as a result of
  noise.  Such a stringent optical non-detection criterion is required
  to keep the contamination low.\label{fig:seleff}}
\end{figure}

\begin{figure}
\epsscale{0.85}
\plotone{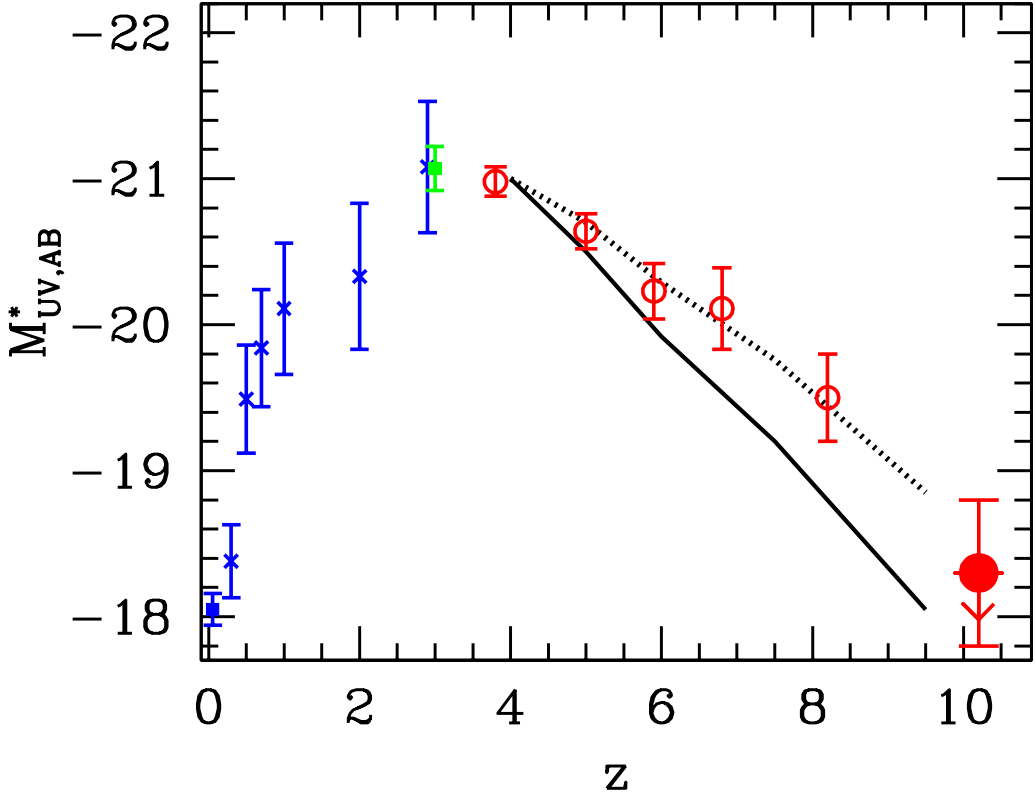}
\caption{$J_{125}$-dropout search results expressed as constraints on
  the characteristic luminosity $M_{UV,AB} ^{*}$ at $z\sim10$.
  $\phi^*$ is taken to be 0.0012 Mpc$^{-3}$ for the $z\sim10$ fit
  (similar to the values found at $z\sim3$-8$^{1,22}$) while $\alpha$
  is fixed to $-1.74$ -- the same value as found at $z\sim6$.$^1$ The
  $z\sim10$ constraint is shown as a solid circle based upon an
  estimate of 0.8 probable $J_{125}$-dropout candidates and as an
  upper limit (assuming none of our candidates correspond to $z\sim10$
  galaxies).  These are essentially identical given our single
  $z\sim10$ candidate.  Also shown are the characteristic luminosities
  estimated for the UV LF at $z\lesssim8$.$^{1,2,3,22,50}$ The solid
  black line shows the evolution predicted from the halo mass
  function$^{51}$ assuming a mass-to-light ratio which is constant
  while the dashed black line shows this evolution assuming a
  mass-to-light ratio which varies as $(1+z)^{-1}$.  These lines are
  identical to those in Figure 10 of Bouwens et al.$^{22}$.  While it
  is unclear whether the $UV$ LF at $z\sim10$ is well represented by a
  Schechter function and whether $\phi^*$ and $\alpha$ remain
  relatively constant over this redshift interval, the characteristic
  luminosity $M^*$ seems to be a convenient one-parameter way of
  characterizing the evolution of the UV LF over a wide redshift
  range.  From this figure, it is clear that observed evolution in the
  $UV$ LF reported from $z\sim8$ to $z\sim4$ continues out to
  $z\sim10$.\label{fig:schlf}}
\end{figure}

\begin{deluxetable}{lcccc}
\tablewidth{14.5cm}
\tabletypesize{\footnotesize}
\tablecaption{$UV$ Luminosity Densities and Star Formation Rate Densities.\tablenotemark{a}\label{tab:sfrdens}}
\tablehead{
\colhead{} & \colhead{} & \colhead{$\textrm{log}_{10} \mathcal{L}$} & \multicolumn{2}{c}{$\textrm{log}_{10}$ SFR density} \\
\colhead{} & \colhead{} & \colhead{(ergs s$^{-1}$} & \multicolumn{2}{c}{($M_{\odot}$ Mpc$^{-3}$ yr$^{-1}$)} \\
\colhead{Dropout Sample} & \colhead{$<z>$} & \colhead{Hz$^{-1}$ Mpc$^{-3}$)} & \colhead{Uncorrected} & \colhead{Corrected\tablenotemark{b}}}
\startdata
$z$$^9$ & 6.8 & 25.80$\pm$0.10 & $-2.10\pm0.10$ & $-2.10\pm0.10$ \\
$Y$$^9$ & 8.0 & 25.57$\pm$0.11 & $-2.33\pm0.11$ & $-2.33\pm0.11$ \\
$J$ (This work: 0.8 candidates) & 10.3 & 24.29$_{-0.76}^{+0.51}$ & $-$3.61$_{-0.76}^{+0.51}$ & $-$3.61$_{-0.76}^{+0.51}$\\
$J$ (This work: no candidates) & 10.3 & $<$24.42\tablenotemark{c} & $<-$3.48\tablenotemark{c} & $<-$3.48\tablenotemark{c}\\
\enddata
\tablenotetext{a}{Integrated down to $\sim$$-17.9$ AB mag.}\
\tablenotetext{b}{The blue $UV$-continuum slopes $\beta\lesssim-2.5$
  observed at $z\sim7$-8 suggest very little dust
  extinction.$^{2,3,49}$ As a result, the corrected and uncorrected
  SFR densities are the same.}
\tablenotetext{c}{Upper limits here are $1\sigma$ (68\% confidence).}
\end{deluxetable}

\section{Implications for the $UV$ LF at $z\sim10$} 

The depth and area of the present WFC3/IR observations have provided
us with our best opportunity yet to find star-forming galaxies at
$z\sim10$, and therefore it is not surprising that we were able to
find one plausible $z$$\sim$10 $J_{125}$-dropout candidate.

The present search results permit us to significantly improve the
constraints on the LF at $z$$\sim$10.  We begin by exploring stepwise
constraints on the UV LF.  The simulations described in \S9 are used
to estimate the effective search volume as a function of absolute
magnitude.  Both the wide-area ERS and ultra-deep HUDF searches are
included in these effective volume estimates.  These volumes
implicitly include the effects of incompleteness and are an integral
of the selection efficiencies with redshift
(Figure~S\ref{fig:seleff}).  The LF is then equal to the number of
$z\sim10$ candidates found divided by these effective volumes.  Based
upon our contamination estimates (\S7), we assume that 0.8
$J_{125}$-dropout candidates in our HUDF09 selection are probable
$z\sim10$ galaxies, but also consider the possibility that none
correspond to $z\sim10$ galaxies.  Our stepwise LFs are presented in
Figure 3 of the main text.

It is also interesting to examine the present LF results in terms of a
Schechter parametrization, given the utility such parametrizations
have had in modelling the LF over a wide range in redshifts (though we
recognize that this parameterization may not be appropriate at
$z\sim10$).  Again we consider two cases, first that 0.8
$J_{125}$-dropout candidates from our HUDF09 selection correspond to
$z\sim10$ galaxies and second that no candidates do.  To quantify the
constraints we can set on the Schechter parameters at $z\sim10$, we
calculate the number of $J_{125}$-dropouts we would expect to find to
these magnitude limits, using the selection efficiencies we can infer
from the simulation results in \S9.  For simplicity, we consider the
constraints on the characteristic luminosity $M^*$ at $z\sim10$, if
the normalization $\phi^*$ and faint-end slope $\alpha$ is equal to
$\sim$0.0012 Mpc$^{-3}$ and $\sim-1.74$ -- similar to that found at
lower redshift $z\sim4-6$.$^{1,22}$ In detail, we find that
$M_{UV}^{*} = -18.3\pm0.5$ AB mag at $z\sim10$, assuming that 0.8
$J_{125}$-dropout candidates from the present sample are at $z\sim10$.
In the case that none are, $M_{UV}^{*} > -18.3$ AB mag at $z\sim10$,
which is $\gtrsim$2.5 mag fainter than the $M_{UV,AB}^{*}\sim-21$ AB
mag derived$^{1,28}$ at $z\sim4$ and the
$M_{UV,AB}^{*}\sim-19.5\pm0.3$ estimated$^3$ at $z\sim8$ --
illustrating how dramatically the $UV$ LF is evolving at high
redshift.  While we have no evidence that the UV LF maintains a
Schechter parametrization out to $z\sim10$ with a relatively constant
$\phi^*$ and $\alpha$, parametrizing the evolution in terms of the
characteristic luminosity $M^*$ does provide us with a convenient way
of quantifying the overall rate of evolution in the $UV$ LF.

\section{Constraints on the $UV$ Luminosity Density/SFR density at
  $z\sim10$} 

The present $J_{125}$-dropout searches permit us to set constraints on
the luminosity density at $z\sim10$ brightward of 0.06 $L_{z=3}^{*}$.
To derive these constraints, we integrate up the LF results from \S10
to the survey limits.  Again, we consider the case that 0.8
$J_{125}$-dropout candidates from our selection correspond to
$z\sim10$ galaxies and the case that none do.  The derived $UV$
luminosity densities are then converted into the equivalent SFR
densities, adopting the Madau et al.$^{52}$ conversion factor.  The
Madau et al.$^{52}$ conversion factor assumes a constant rate of star
formation for $\gtrsim100$ Myr. This suggests that it may be
inappropriate to use this conversion at $z\sim10$.  Interestingly, the
time scale inferred for star-formation in star-forming galaxies at
$z$$\sim$7-8 is nonetheless quite extended,$^{12,13}$ with inferred
ages of $\sim$300-400 Myr.  This suggests that even at $z\sim10$ it
may not be unreasonable to suppose that galaxies experience extended
star-formation for $\gtrsim$100 Myr.  However, we recognize that the
Madau et al.$^{52}$ approach, while convenient, is limited, and look
forward to more sophisticated analyses in the future.$^{53}$

The results are presented in Table~S\ref{tab:sfrdens} and Figure 4 of
the main text, alongside our most recent results at $z\sim7$ and
$z\sim8$.$^9$ Since essentally all galaxies in the luminosity range of
our samples (i.e., $\gtrsim-19.5$ AB mag) show very blue
$UV$-continuum slopes $\beta$ ($\beta\lesssim-2.5$) at
$z\gtrsim5$,$^{25,49}$ no dust obscuration is assumed in the SFR
density estimates at high redshift.

\clearpage

\section{References}

\noindent
31.  Magee, D.~K., Bouwens, R.~J., \& Illingworth, G.~D.\ NICRED: A
NICMOS Image Processing Pipeline, \textit{Astronomical Data Analysis
  Software and Systems XVI}, \textbf{376}, 261 (2007).\vspace{0.3cm}
\\
32.  Koekemoer, A.~M., Fruchter, A.~S., Hook, R.~N., \& Hack,
W.\ MultiDrizzle: An Integrated Pyraf Script for Registering, Cleaning
and Combining Images, \textit{The 2002 HST Calibration Workshop :
  Hubble after the Installation of the ACS and the NICMOS Cooling
  System}, 337 (2002).\vspace{0.3cm} \\
33.  Giavalisco, M., et al.\ The Great Observatories Origins Deep
Survey: Initial Results from Optical and Near-Infrared Imaging,
\textit{Astrophys. J. Lett.}, \textbf{600}, 93-98
(2004).\vspace{0.3cm} \\
34.  Riess, A.~G., et al.\ New Hubble Space Telescope Discoveries of
Type Ia Supernovae at $z\geq 1$: Narrowing Constraints on the Early
Behavior of Dark Energy, \textit{Astrophys. J.}, \textbf{659}, 98-121
(2007).\vspace{0.3cm} \\
35.  Bouwens, R.J., Illingworth, G.D., Blakeslee, J.P., \& Franx, M.
Galaxies at z$\sim$6: The UV Luminosity Function and Luminosity
Density from 506 UDF, UDF-Ps, and GOODS i-dropouts,
\textit{Astrophys. J.}, \textbf{653}, 53-85 (2006).\vspace{0.3cm} \\
36.  Bouwens, R.J., Broadhurst, T.J., Illingworth, G.D.  Cloning
Dropouts: Implications for Galaxy Evolution at High Redshift,
\textit{Astrophys. J.} \textbf{593}, 640-660 (2003).\vspace{0.3cm} \\
37.  Bertin, E.\ \& Arnouts, S.\ SExtractor: Software for source
extraction, \textit{Astron. Astrophys. Suppl.}, \textbf{117}, 393-404
(1996).\vspace{0.3cm} \\
38.  Kron, R. G. Photometry of a complete sample of faint galaxies,
\textit{Astrophys. J. Suppl.}, \textbf{43}, 305-325
(1980).\vspace{0.3cm} \\
39.  Szalay, A.~S., Connolly, A.~J., \& Szokoly, G.~P.\ Simultaneous
Multicolor Detection of Faint Galaxies in the Hubble Deep Field,
\textit{Astron. J.}, \textbf{117}, 68-74 (1999).\vspace{0.3cm} \\
40.  Labb\'{e}, I., Bouwens, R., Illingworth, G.~D., \& Franx,
M.\ Spitzer IRAC Confirmation of $z_{850}$-Dropout Galaxies in the
Hubble Ultra Deep Field: Stellar Masses and Ages at $z\sim7$,
\textit{Astrophys. J. Lett.}, \textbf{649}, 67-70 (2006).\vspace{0.3cm} \\
41.  Coleman, G.~D., Wu, C.-C., \& Weedman, D.~W.\ Colors and
magnitudes predicted for high redshift galaxies,
\textit{Astrophys. J. Suppl.}, \textbf{43}, 393-416 (1980).\vspace{0.3cm} \\
42.  Knapp, G.~R., et al.\ Near-Infrared Photometry and Spectroscopy
of L and T Dwarfs: The Effects of Temperature, Clouds, and Gravity,
\textit{Astron. J.}, \textbf{127}, 3553-3578 (2004).\vspace{0.3cm} \\
43.  Oesch, P.~A., et al.\ The UDF05 Follow-Up of the Hubble Ultra
Deep Field. II. Constraints on Reionization from Z-Dropout Galaxies,
\textit{Astrophys. J.}, \textbf{690}, 1350-1357 (2009).\vspace{0.3cm}
\\
44.  Whitelock, P., et al.\ Mass-losing stars in the South Galactic
CAP, \textit{Month. Not. R. Astron. Soc.}, \textbf{276}, 219-254
(1995).\vspace{0.3cm} \\
45.  Dickinson, M., et al.\ The Unusual Infrared Object HDF-N
J123656.3+621322, \textit{Astrophys. J.}, \textbf{531}, 624-634
(2000).\vspace{0.3cm} \\
46.  Bouwens, R.~J., Illingworth, G.~D., Blakeslee, J.~P., Broadhurst,
T.~J., \& Franx, M.\ Galaxy Size Evolution at High Redshift and
Surface Brightness Selection Effects: Constraints from the Hubble
Ultra Deep Field, \textit{Astrophys. J. Lett.}, \textbf{611}, 1-4
(2004).\vspace{0.3cm} \\
47.  Ferguson, H.~C.~et al.\ The Size Evolution of High-Redshift
Galaxies, \textit{Astrophys. J. Lett.}, \textbf{600}, 107-110
(2004).\vspace{0.3cm} \\
48.  Oesch, P.A., et al.\ Structure and Morphologies of z$\sim$7-8
Galaxies from Ultra-deep WFC3/IR Imaging of the Hubble Ultra-deep
Field, \textit{Astrophys. J.}, \textbf{709}, 21-25 (2010).\vspace{0.3cm} \\
49.  Bouwens, R.~J., et al.\ Very Blue UV-Continuum Slope $\beta$ of
Low Luminosity $z\sim7$ Galaxies from WFC3/IR: Evidence for Extremely
Low Metallicities?, \textit{Astrophys. J. Lett.}, \textbf{708}, 69-73
(2010).\vspace{0.3cm} \\
50.  Arnouts, S., et al.\ The GALEX VIMOS-VLT Deep Survey Measurement
of the Evolution of the 1500\AA$~$Luminosity Function,
\textit{Astrophys. J. Lett.}, \textbf{619}, 43-46 (2005).\vspace{0.3cm} \\
51.  Sheth, R.~K.~\& Tormen, G.\ Large-scale bias and the peak
background split, \textit{Month. Not. R. Astron. Soc.}, \textbf{308},
119-126 (1999).\vspace{0.3cm} \\
52.  Madau, P., Pozzetti, L. \& Dickinson, M. The Star Formation
History of Field Galaxies, \textit{Astrophys. J.}, \textbf{498},
106-116 (1998).\vspace{0.3cm} \\
53.  Verma, A., Lehnert, M.~D., F{\"o}rster Schreiber, N.~M., Bremer,
M.~N., \& Douglas, L.\ Young Galaxies in the Early Universe: The
Physical Properties of Luminous $z\sim5$ LBGs Derived from Their
Rest-frame UV to Visible SEDs, \textit{Month. Not. R. Astron. Soc.},
\textbf{377}, 1024-1042 (2007).\vspace{0.3cm} \\

\begin{figure}
\epsscale{0.85}
\plotone{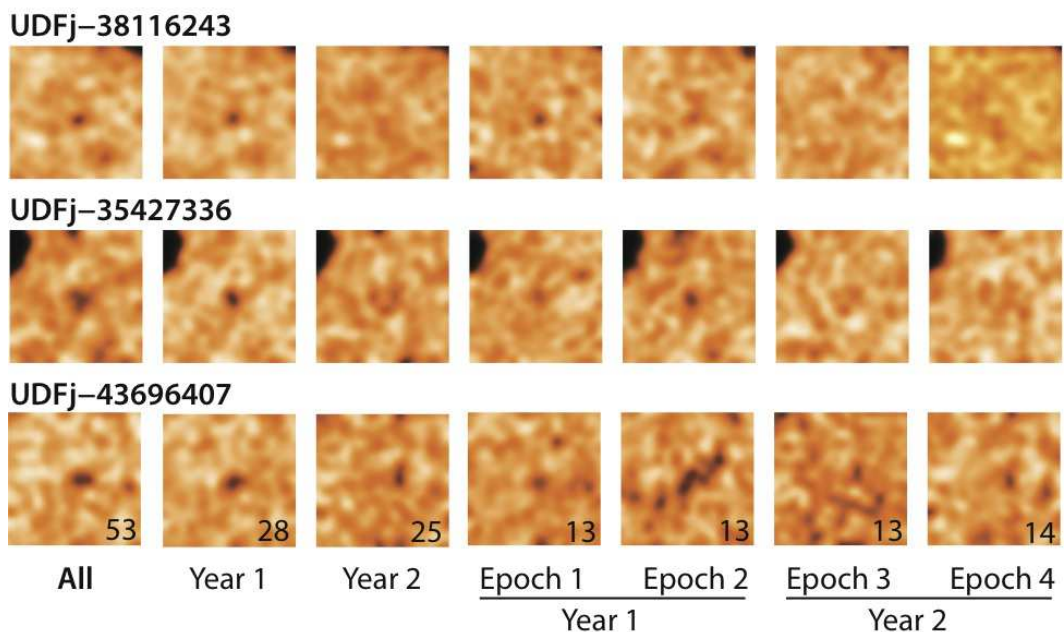}
\caption{Appearance of the $z\sim10$ $J_{125}$-dropout candidates we
  had identified in the first-year data in different subsets of the
  full data set.  Shown is their appearance in the full data set, in
  the first and second year data, and in 13-orbit subsets of the
  ultra-deep WFC3/IR $H_{160}$-band observations over the HUDF (53
  orbits in total).  The number of orbits of observations for each
  subset is given in the lower right corner.  While these candidates
  are visible at modest significance in the two 50\% splits of the
  first-year data, they are not visible at much significance in the
  two 50\% splits of the second-year data (and hence cannot be
  considered to be robust candidates in the full data set: see
  Appendix A).\label{fig:epochold}}
\end{figure}

\appendix

\section{Assessment of our First-year $J_{125}$ Dropout Selection}

Shortly after the first-year of ultra-deep WFC3/IR observations over
the HUDF were obtained, we carried out a search for $z\sim10$
$J_{125}$-dropout candidates.  Three $J_{125}$-dropout candidates were
identified.  These candidates showed $5\sigma$ detections in scalable
elliptical apertures and had measured $J_{125}-H_{160}$ colors $>$1.2.
They showed no significant detection in the optical data.  These three
$J_{125}$-dropout candidates also showed no significant detection in
the IRAC $3.6\mu$ and $4.5\mu$ observations -- consistent with the
much shallower depths of the IRAC observations.  These candidates
plausibly corresponded to star-forming galaxies at $z\sim10$.  See
Bouwens et al. arXiv:0912.4263 for details.

The availability of a second year of ultra-deep WFC3/IR observations
over the HUDF (25 orbits in the $H_{160}$ band) provided us with the
opportunity to test the robustness of these candidates.  If real, one
would have expected the candidates to be detected at $\geq$3.5$\sigma$
in the new observations.  Figure~S\ref{fig:epochold} shows the new
observations we have for our three candidates, the first-year
observations and the full two-year observations considered together.
Strikingly, two of the candidates are not even significant at
$1\sigma$ in the new observations, which suggests two candidates from
the first-year observations were simply spurious.  For the third
candidate UDFj-38116243, a 1.3$\sigma$ detection is found at the
position of the source.  While this is consistent with the candidate
UDFj-38116243 corresponding to an actual source (with an apparent
magnitude at least 1 mag fainter than what we estimated from the
first-year observations), we no longer consider UDFj-38116243 to be a
viable $z\sim10$ candidate. The large discrepancy between the
5.6$\sigma$ detection in the first year observations and the
1.3$\sigma$ detection in the second year observations gives us little
confidence that UDFj-38116243 is a real source and suggests that it
may be spurious.

This is surprising, given the nominal detection of the three
candidates in the first year observations at $\sim$5$\sigma$, the
visibility of the candidates in 50\% subsets of that first-year data,
and the apparent detection of two of the candidates in the NICMOS
$H_{160}$-band data.  Nonetheless, the fact that all three candidates
were identified close to the $5\sigma$ completeness limit of the
first-year observations -- where the contamination increases
relatively rapidly as the S/N decreases further -- indicated that the
candidates were not the most robust (as we noted).

How is it possible that apparent $5\sigma$ detections might not prove
to be robust?  A primary reason is the way in which the apertures were
defined.  In searching for possible $5\sigma$ detections in the
first-year observations, we utilized scalable Kron-style apertures
(Kron [1980] factor of 1.2).  Such apertures -- while working very
well to measure the fluxes for faint sources -- can lead to biased
measurements of source significance.  This is because the apertures
that are used to evaluate the significance of the sources are also
established by the same apparent flux distribution.  This can lead to
overestimates of the significance by as much as $0.5\sigma$ to
$1.0\sigma$ for $\sim$4$\sigma$ sources, and so some sources that are
really $<5\sigma$ can appear at $>5\sigma$.  Even for purely Gaussian
noise, one would expect a modest number $\gtrsim$3 of purely spurious
$\sim$4$\sigma$ candidates.

A second contributing factor is our slight underestimate of the noise
in the first-year observations.  The extent of the underestimate
appears to have been just 5\%, as obtained from our current estimates
of the noise where we actually fit to the full noise distribution
(i.e., Figure~S\ref{fig:noise}), but this is large enough to make a
$4.7\sigma$ fluctuation look like a $5\sigma$ fluctuation.

Third, as expected, there appears to have been a very slight
non-Gaussianity to the noise distribution.  While it is remarkably
Gaussian to large $\sigma$ (Figure~S\ref{fig:noise}), the small
non-Gaussianity present could possibly have enhanced the number of
spurious sources with apparent $\sim$4-5$\sigma$ significance levels
over that expected from Gaussian statistics by 50\%.  The factor is
quite poorly determined, however, since it is quite challenging to
establish deviations from Gaussianity at $\sim$4$\sigma$.  The
existence of a $\sim$5$\sigma$ single-band detection in the 34-orbit
$J_{125}$-band image (see \S7) provides some evidence that
non-Gaussianity could play a role, but we emphasize that such effects
are likely to be less important in the full two-year 53-orbit
$H_{160}$-band data (which consisting of a larger number of exposures
should have more Gaussian noise properties).

Fourth, we now realize that the tests we used for establishing the
level of contamination from spurious sources led to an underestimate
of the prevalence of those sources. This is likely to be a significant
issue with our earlier analysis. We have realized that our previous
simulations likely suffered from two shortcomings which caused us to
underestimate the number of spurious sources significantly. We did not
include (1) the expected population of very faint ("below-the-limit")
$\sim$30-33 mag sources in the simulations and (2) large foreground
sources.  The expected population of faint sources (1) would act to
significantly increase the number of positive $\sigma$ fluctuations.
By ignoring this effect, we significantly underpredicted the number of
spurious sources.  (2) The effect of bright sources may not be as
significant, but they would also play a role by making the sky
background more difficult to determine (some empty regions of the
image are on the wings of bright objects and some regions are not).
This would result in significant underestimates of the sky levels for
many sources and hence overestimates of their significance levels.  By
not including bright sources, our simulations would not accounted for
the effect they would on the background estimates, and hence we could
easily have underpredicted the number of spurious sources present in
the real observations.

Taken together these results show that no one factor was dominant in
why the sources proved to be unreliable.  Probably the most
significant factor was that the flux in the candidates was
overestimated by the use of the scalable elliptical apertures (by
5-15\%).  This, combined with the slight underestimate of the noise
level in the data (by 5\%), meant that the S/N of the sources was not
$>$5$\sigma$, but $<$5$\sigma$, and closer to 4$\sigma$.  Since the
likelihood of contamination from spurious sources increases
exponentially as the S/N decreases, the contamination rate was not the
small numbers expected for S/N $>$ 5, but the much larger rate for S/N
$\sim$ 4.  The spurious source contamination rate was likely an order
of magnitude (or more) larger.

In summary, the combination of (1) the sources being fainter than we
derived from the scalable aperture and (2) the noise being
underestimated led to our selecting sources that had lower S/N than
5$\sigma$. For such sources the spurious source contamination rate was
much higher. These factors combined to make it much more likely that
the sources were spurious (in two cases) or, at best, much less
significant (the other case of the three). The unfortunate
conincidence of a marginal signal (and likely spurious event also) in
the NICMOS observations gave us an added sense of security that proved
unwarranted.

\clearpage

\end{document}